# On the origin of critical temperature enhancement in atomically-thin superconductors

E.F. Talantsev[1,*], W.P. Crump[1], J.O. Island[2,**], Ying Xing[3,4,5], Yi Sun[3,4], Jian Wang[3,4], and J.L. Tallon[1,6].

[1]Robinson Research Institute, Victoria University of Wellington, 69 Gracefield Road, Lower Hutt, 5040, New Zealand

[2]Kavli Institute of Nanoscience, Delft University of Technology, Lorentzweg 1, Delft 2628 CJ, The Netherlands

[3]International Center for Quantum Materials, School of Physics, Peking University, Beijing 100871, People's Republic of China

[4]Collaborative Innovation Center of Quantum Matter, Beijing, People's Republic of China

[5]Beijing Key Laboratory of Optical Detection Technology for Oil and Gas, China University of Petroleum, Beijing 102249, People's Republic of China

[6]MacDiarmid Institute for Advanced Materials and Nanotechnology, P.O. Box 33436, Lower Hutt 5046, New Zealand

*Corresponding author: evgeny.talantsev@vuw.ac.nz

**Present address: Department of Physics, University of California, Santa Barbara, CA 93106, USA

*Abstract*

Recent experiments showed that thinning gallium, iron selenide and 2H tantalum disulfide to single/several monoatomic layer(s) enhances their superconducting critical temperatures. Here, we characterize these superconductors by extracting the absolute values of the London penetration depth, the superconducting energy gap, and the relative jump in specific heat at the transition temperature from their self-field critical currents. Our central finding is that the enhancement in transition temperature for these materials arises from the opening of an additional superconducting gap, while retaining a largely unchanged "bulk" superconducting gap. Literature data reveals that ultrathin niobium films similarly develop a second superconducting gap. Based on the available data, it seems that, for type-II superconductors, a new superconducting band appears when the film thickness becomes smaller than the out-of-plane coherence length. The same mechanism may also be the cause of enhanced interface superconductivity.

1. Introduction

Fundamental mechanisms governing superconductivity in the two-dimensional (2D) limit represent a long-standing problem in physics [1-5]. The conventional picture is that the reduction in dimensionality causes the growth of fluctuations and a weakening of superconductivity [6,7] such that gradual sample thinning (or reduction of cross-sectional dimensions in the case of nanowires) causes a superconductor-to-insulator [6,8] or superconductor-to-normal metal [6,9] transition. The current status of the subject was recently reviewed in Ref. 10.

On the other hand, 2D systems will often exhibit a van Hove singularity with an associated divergence in the electronic density of states (DOS) which can, within a BCS scenario, result in an enhanced superconducting transition temperature, $T_c$ [11]. This is particularly relevant in the case of the cuprate high-$T_c$ superconductors where a saddle point singularity lies close to the Fermi level and has a clear signature in the evolution of $T_c$ with doping [12]. In low-dimensional systems there is therefore a tension between the twin roles of fluctuations and an enhanced DOS. As to which wins remains a question of detail. In this context, recent studies of single-atomic-layer films of FeSe [13,14], double- and triple-atomic-layer of hexagonal gallium films [15,16], and several-atomic-layer exfoliated films of 2H-TaS$_2$ [17] showed that, despite reduction in film thickness, the transition temperature increases remarkably. In each case the explanation was proposed that $T_c$ rises due to an enhancement in the effective electron–phonon coupling constant [13-17]. The current status of studies of the FeSe single-atomic layer superconductor was recently reviewed in Ref. [18].

In this paper we analyze the experimental self-field critical current density, $J_c(sf,T)$, of ultra-thin films including: single-atomic-layer FeSe, few-atomic-layer hexagonal Ga, Mo$_2$C, and exfoliated 2H-TaS$_2$, 2H-NbSe$_2$, and 2H-MoS$_2$ to extract their fundamental superconducting parameters. Some of these systems exhibit a significantly enhanced $T_c$ over that observed in the bulk state, and in most cases their superconducting parameters were not previously established for such ultra-thin films. Because $J_c(sf,T)$ is directly related to the London penetration depth, $\lambda(T)$, [19] we are able to fit the data using modified BCS-like equations, as applicable for single- or multi-band superconductors and weak- or strong-coupling superconductors. We have made the fitting procedures, which are quite complex, available online for public use [20]. These yield, as fit parameters, values for the ground-state London penetration depth, $\lambda(0)$, $T_c$, the ground-state superconducting energy gap, $\Delta(0)$, and the jump in electronic specific heat $\Delta C/C$ at $T_c$ for each band. In all investigated atomically thin superconductors for which the enhancement of $T_c$ was observed, we find that the enhancement is always associated with the opening of an additional larger gap while the (smaller) bulk gap remains essentially unchanged as the sample is thinned towards the 2D limit. We infer from this that the enhancement in $T_c$ is therefore not primarily associated with enhanced coupling, or an increased energy scale for the pairing boson, but arises from additional gapping on the Fermi surface(s). Significantly, this additional gap seems to open when the ground-state amplitude of the out-of-plane coherence length, $\xi_c$, exceeds the film thickness.

## 2. Model description

If a superconductor has rectangular cross-section then the experimentally-measured critical current $I_c$ can be converted to a critical current density $J_c = I_c/(4ab)$, where, in accordance with commonly-accepted convention [21-23], $2a$ is the width, and $2b$ is the thickness of the conductor. (These definitions arise from the conveniently chosen axes for considering Meissner currents in rectangular superconductors, where the sample width lies along the X axis and the sample thickness along the Y axis. Because the solution to the London equations for the field in a rectangular film involves a hyperbolic sine function, $\sinh(y/\lambda)$, it is convenient for $y$ to run from $-b$ to $+b$, so that the thickness is $2b$. Similarly the width runs from $x = -a$ to $+a$ so that the width is $2a$).

Recently we showed [19] that in thin film superconductors with thicknesses less than the London penetration depth (which is the case for all films we consider herein) the self-field critical current is reached when the critical current density, $J_c(\text{sf})$, reaches $B_c/(\mu_0\lambda)$ for type I superconductors or $B_{c1}/(\mu_0\lambda)$ for type II superconductors. Here $B_c$ is the thermodynamic critical field, $B_{c1}$ is the lower critical field and $\lambda$ is the London penetration depth. Thus [19]:

$$J_c(\text{sf},T) = \frac{\phi_0}{2\sqrt{2}\,\pi\,\mu_0} \cdot \frac{\kappa}{\lambda^3(T)} \tag{3}$$

for type-I superconductors, and

$$J_c(\text{sf},T) = \frac{\phi_0}{4\pi\,\mu_0} \cdot \frac{(\ln(\kappa)+0.5)}{\lambda^3(T)} \tag{4}$$

for type-II superconductors, where, $\mu_0$ is the magnetic permeability of free space, $\phi_0$ is flux quantum, and $\kappa = \lambda/\xi$ is the Ginsburg-Landau parameter (any temperature dependence of which we neglect). By measuring $J_c(\text{sf},T)$ and knowing the magnitude of $\kappa$ the inversion of Eq. 4 gives us a tool to convert $J_c(\text{sf},T)$ to absolute values of $\lambda(T)$. Figure 1(a) illustrates this method, where we used the $J_c(\text{sf},T)$ data from Clem et al. for a NbN film ($2a = 6.0$ μm, $2b = 22.5$ nm) [24] where $\kappa = 40$ for NbN [25].

If $J_c(\text{sf},T)$ measurements are performed to low enough temperature, by which conventional agreement is $T < T_c/3$ [26], then the absolute magnitudes of the ground-state superconducting energy gap, $\Delta(0)$, and London penetration depth, $\lambda(0)$ may be deduced from a data fit to the low-temperature asymptotes of the Bardeen-Cooper-Schrieffer (BCS) theory [27]:

$$\lambda(T) = \frac{\lambda(0)}{\sqrt{1 - 2\sqrt{\frac{\pi \Delta(0)}{k_B T}} \cdot e^{-\frac{\Delta(0)}{k_B T}}}} \tag{5}$$

for *s*-wave [28], and:

$$\lambda(T) = \frac{\lambda(0)}{\sqrt{1 - 2\frac{k_B T}{\Delta_m(0)}}} \tag{6}$$

for *d*-wave [29], where $\Delta_m$ is the amplitude of the *k*-dependent *d*-wave gap, $\Delta = \Delta_m \cos(2\theta)$.

Based on Eqs. 5 and 6, we can conclude that the $J_c(\text{sf},T)$ of *s*-wave superconductors is exponentially flat for $T < T_c/4$:

$$J_c(\text{sf},T) \approx \frac{\phi_0}{4\pi \mu_0} \cdot \frac{(\ln(\kappa) + 0.5)}{\lambda^3(0)} \cdot \left(1 - 3\sqrt{\frac{\pi \Delta(0)}{k_B T}} \cdot e^{-\frac{\Delta(0)}{k_B T}}\right), \tag{7}$$

while at the same conditions $J_c(\text{sf},T)$ of *d*-wave superconductors:

$$J_c(\text{sf},T) \approx \frac{\phi_0}{4\pi \mu_0} \cdot \frac{(\ln(\kappa) + 0.5)}{\lambda^3(0)} \cdot \left(1 - 3\frac{k_B T}{\Delta_m(0)}\right) \tag{8}$$

is a linear function with the slope inversely proportional to $\Delta(0)$.

For the above case of NbN [24], which is an *s*-wave superconductor, the corresponding fits are presented in Fig. 1(b). The fit quality was assessed by the goodness of fit parameter, $R$, and coefficients of mutual dependency of fitting parameters. These were calculated in the same manner for all samples analyzed in this paper. Details for the procedure are presented in Supplementary Information (SI). The derived ground-state London penetration depth, $\lambda(0) = 194.1 \pm 0.1$ nm, is in remarkable agreement with the independently measured value, $\lambda(0) = 194$ nm, for NbN [30] (see green data point on the *y*-axis of Fig. 1(b)). Also, the deduced $\Delta(0) = 2.12 \pm 0.01$ meV, and BCS ratio of $2\Delta(0)/k_BT_c = 4.17 \pm 0.02$, confirm the strong-coupling scenario for NbN and well match reported measurements of $2\Delta(0)/k_BT_c = 4.25$ [31].

Previously, we performed this analysis for a wide range of thin-film superconductors, including metals, nitrides, oxides, cuprates, pnictides, borocarbides, $MgB_2$ and heavy Fermions [19]. Derived values of $\lambda(0)$ and $\Delta(0)$ were in very good agreement with values measured by conventional (and usually much more complex) techniques [19].

Recently [32], other authors used our approach (Eq. 4) to construct a Uemura plot [33] for ionic-liquid gated $YBa_2Cu_3O_{7-x}$ films of thickness just 4-5 nm. The results nicely concurred with the Uemura plots obtained using other techniques such as muon spin relaxation. In the case of the highly compressed sulfur hydride superconductor, $H_3S$, with record transition temperature of 203 K [34], our approach (Eqs. 5 and 7) is perhaps the only currently available technique to derive the magnitude of superconducting energy gap for this material, $\Delta(0) = 27.8 \pm 0.2$ meV [35].

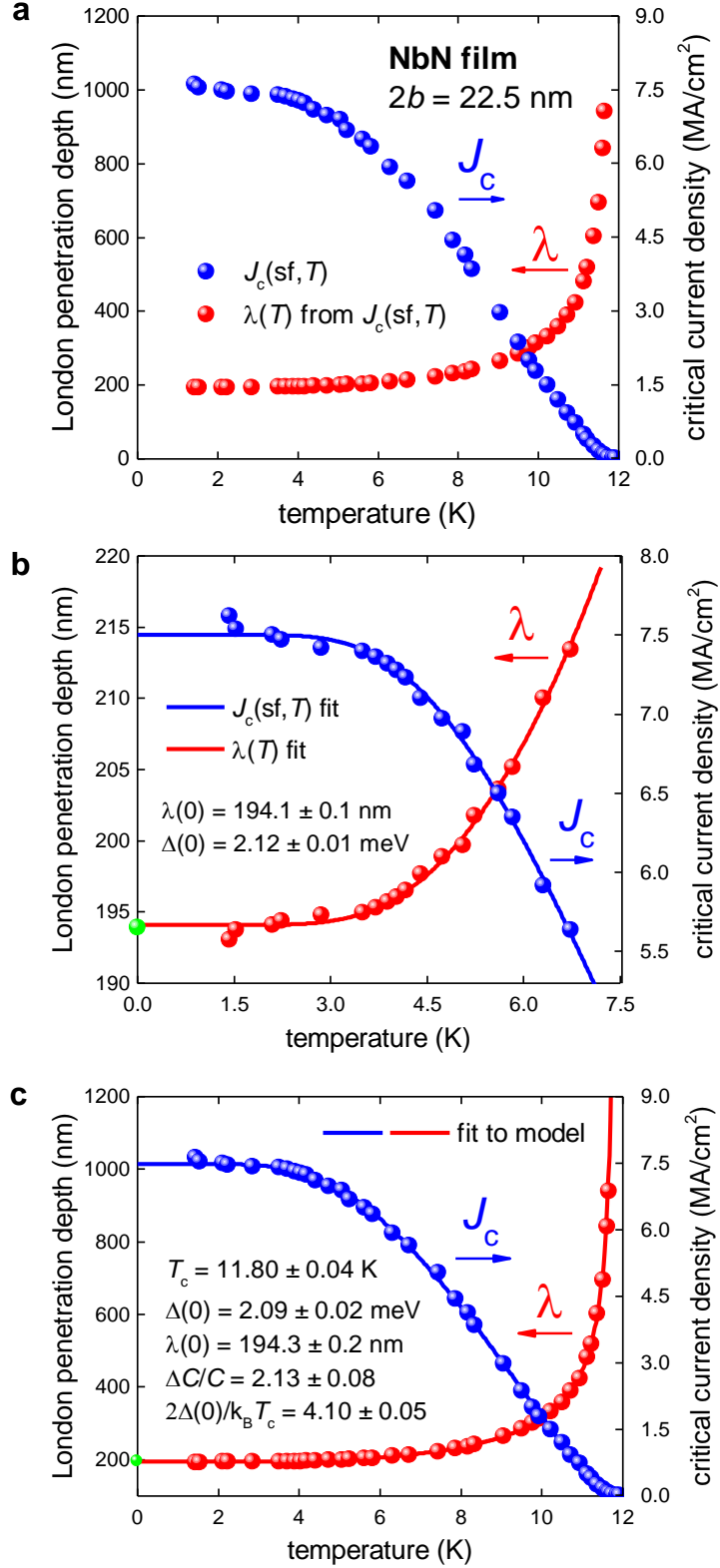

**Figure 1.** Experimental $J_c(\text{sf},T)$ data and fits for a thin ($2b = 22.5$ nm) film of NbN (right axis, blue) together with values of $\lambda(T)$ (left axis, red) derived by Eq. 4. The single green data point at $T = 0$ K is the independently-reported ground-state value of $\lambda(0) = 194$ nm [25]. (a) Full scale picture; (b) Magnified plot of low-temperature region. The solid curves are fits to low-temperature BCS asymptotes, Eqs. 5 and 7. Derived $2\Delta(0)/k_B T_c = 4.17 \pm 0.02$. Fit quality

is $R = 0.9953$. (c) The solid curves are the BCS-like fits using Eqs. 10 and 11. Fit quality is $R = 0.9988$.

In this paper, we employ the general approach of BCS theory [27], in which the thermodynamic properties of a superconductor are derived from the superconducting energy gap, $\Delta(T)$. We use the temperature-dependent superconducting gap $\Delta(T)$ equation given by Gross [36] (which allows variation in the coupling strength):

$$\Delta(T) = \Delta(0) \cdot \tanh\left(\frac{\pi k_B T_c}{\Delta(0)} \cdot \sqrt{\eta\left(\frac{\Delta C}{C}\right)\left(\frac{T_c}{T} - 1\right)}\right) \qquad (9)$$

where $\Delta C/C$ is the relative jump in electronic specific heat at $T_c$, and $\eta = 2/3$ for $s$-wave superconductors [35] and $\eta = 7/5$ for $d$-wave superconductors [37]. From this the London penetration depth, $\lambda(T)$ of a flat-band $s$-wave superconductor may be calculated using the BCS expression [27]:

$$\lambda(T) = \frac{\lambda(0)}{\sqrt{1 - \frac{1}{2k_B T} \cdot \int_0^\infty \frac{d\varepsilon}{\cosh^2\left(\frac{\sqrt{\varepsilon^2 + \Delta^2(T)}}{2k_B T}\right)}}} \qquad (10)$$

where, $k_B$ is Boltzmann's constant. By substituting Eq. 10 in Eq. 4:

$$J_c(\text{sf}, T) = \frac{\phi_0}{4\pi\mu_0} \cdot \frac{(\ln(\kappa) + 0.5)}{\lambda^3(0)} \cdot \left(1 - \frac{1}{2k_B T} \cdot \int_0^\infty \frac{d\varepsilon}{\cosh^2\left(\frac{\sqrt{\varepsilon^2 + \Delta^2(T)}}{2k_B T}\right)}\right)^{1.5} \qquad (11)$$

one can fit experimental $J_c(\text{sf},T)$ data to deduce $\lambda(0)$, $\Delta(0)$, $\Delta C/C$ and $T_c$ as free-fitting parameters. The corresponding equation for $d$-wave superconductors can be found elsewhere [37]. To help experimentalists to use our BCS-based model to infer $\lambda(0)$, $\Delta(0)$, $\Delta C/C$ and $T_c$ parameters from measured $J_c$(sf,$T$) data (which is not a trivial mathematical task), we placed our MatLab code for free-online use [20].

Now we illustrate the method using the same $J_c$(sf,$T$) data of Clem et al. [24] and show the results in Fig. 1(c). The fit to the experimental data is excellent ($R = 0.9988$), and the derived fit value $\lambda(0) = 194.3 \pm 0.2$ nm is also in remarkable agreement with the independent measurement of the London penetration depth $\lambda(0) = 194$ nm in NbN [29] (see green data point on the $y$-axis in Fig. 1(c)). This fit also validates our model in terms of its applicability to strong-coupled superconductors, because the derived BCS ratio $2\Delta(0)/k_BT_c = 4.10 \pm 0.05$ and $\Delta C/C = 2.13 \pm 0.08$ confirm the strong-coupling scenario for NbN. Our deduced values are in excellent agreement with the reported measurements for these quantities, $2\Delta(0)/k_BT_c = 4.25$ [31] and $\Delta C/C = 1.90 \pm 0.09$ [38]. Thus, we can conclude that our model adequately derives thermodynamic parameters for strong-coupling superconductors and it does not restrict to just the weak-coupling limit of BCS. More details and examples of the application of this model can be found elsewhere [37].

Next we illustrate the method in the case where the superconductor has two gaps opening on two separate bands as is particularly relevant to the ultra-thin superconductors discussed below. Where there are two strongly-coupled bands, then the so-called $\alpha$-model [39], which utilizes the same common $\lambda(0)$ and $T_c$ values for both bands, can be used:

$$J_c(sf,T)_{total} = \alpha \cdot J_c(sf,T)_{band1} + (1-\alpha) \cdot J_c(sf,T)_{band2}. \tag{12}$$

This means that $J_c$ for each band is calculated using Eq. 11. As a consequence of this, the $J_c$(sf,$T$) dataset should be reasonably rich to derive parameters with acceptable uncertainty. However, dense $J_c$(sf,$T$) data sets are generally unavailable in the literature. Thus, to run the model (Eq. 12) for limited experimental data sets it is convenient to sacrifice some parameter(s) by fixing to certain value(s). For example, because the specific heat jump, $\Delta C_2/C_2$, for the band with the smaller gap is poorly constrained we often fix this to the weak-coupling BCS limit for *s*-wave superconductors, i.e., 1.43 (other authors choose to fix other parameters, see for instance [40]).

Application of the model to an $MgB_2$ thin film ($2b$ = 10 nm) [41] is shown in Fig. 2. The goodness of fit is $R$ = 0.9928 and the derived parameters are again in good agreement with reported values measured by independent techniques, and in particular we deduce $\lambda(0)$ = 85.7 ± 0.2 nm, in good agreement with the independently reported value of 85 nm (green data point on *y*-axis) [42]. More details and examples of the application of this '$\alpha$-model' can be found elsewhere [37].

In the case of a two-band superconductor that has completely decoupled bands, $J_c(sf,T)$ can be written in the form:

$$J_c(sf,T)_{total} = J_c(sf,T)_{band1} + J_c(sf,T)_{band2} \tag{13}$$

where, $J_c(sf,T)$ for each band described by Eq. 11, and indices 1 and 2 will be attributed to separate $\lambda(0)$, $\Delta(0)$, $\Delta C/C$ and $T_c$ values for each band and all eight parameters may be used as free-fitting parameters. Again, we need to note, that for sparse $J_c(sf,T)$ datasets one or two of these eight parameters can be fixed. More details and examples for application of this 'weakly-coupled bands model' can be found elsewhere [37]. The above examples are not simply illustrative of our method but will also be used in the analysis below.

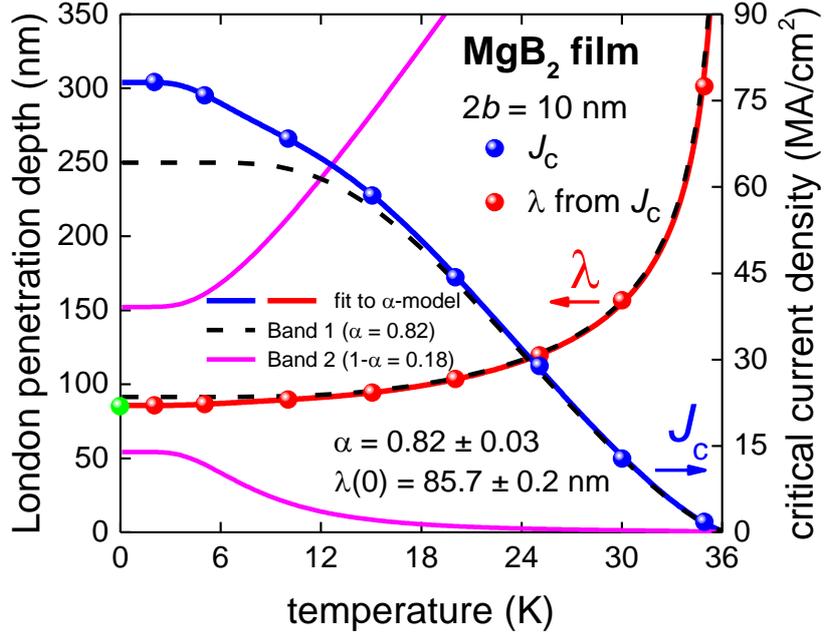

**Figure 2.** Experimental $J_c(sf,T)$ data for a thin ($2b = 10$ nm) film of $MgB_2$ (right axis, blue) together with values of $\lambda(T)$ (left axis, red) derived by Eq. 4. The solid curves are for the 'strongly-coupled bands model' in a BCS-like fit using Eq. 12. The single green data point at $T = 0$ K is the independently-reported ground-state value of $\lambda(0) = 85$ nm [42]. Derived parameters are: $T_c = 36.4 \pm 0.4$ K, $\lambda(0) = 85.7 \pm 0.2$ nm; for band 1: $\Delta_1(0) = 5.6 \pm 0.2$ meV, $\Delta C_1/C_1 = 1.53 \pm 0.15$, for band 2: $\Delta_2(0) = 1.7 \pm 0.2$ meV, $\Delta C_2/C_2 = 1.43$ (fixed). Fit quality is $R = 0.9928$.

### 3. Experimental

Sample fabrication details, critical current measurement techniques, and other characterization methods were reported elsewhere [13-17]. Experimental $J_c(sf,T)$ data sets were not explicitly published in any of these previous publications and we are reporting and analyzing the data herein. To define $J_c$, we use the usual power-law fit [43] of the experimental *I-V* curve by using a voltage criterion of $V = 300$ μV for double-atomic-layer hexagonal Ga and the single-atomic-layer FeSe superconductor, and $V = 5$ μV criterion for all $2H$-$TaS_2$ crystals studied herein.

## 4. Hexagonal double-atomic-layer gallium

Double-atomic-layer hexagonal Ga ($2b = 0.552$ nm) is a type-II superconductor [15] with transition temperature $T_c = 4.5$ K, which is remarkably higher than $T_c = 1.1$ K for bulk Ga which is a type-I superconductor. The crossover from type I to type II reflects the substantial increase in $T_c$ and gap magnitude and the associated reduction in coherence length. Self-field critical currents were measured on a current bridge with width $2a = 2.0$ mm. As the Ginzburg-Landau (GL) parameter $\kappa_c$ for a double-atomic-layer film of hexagonal Ga is unknown, and we cannot use $\kappa$ for bulk Ga, as bulk Ga is a type-I superconductor, we calculated the in-plane coherence length, $\xi_{ab}(0)$, for this film from the reported $R(T, B_\perp)$ data [15]. We applied a criterion of 50% normal resistance recovery to the $R(T, B_\perp)$ curves [15] to define the upper critical field, $B_{c2}(T)$. The fit of $B_{c2}(T)$ data to the GL expression:

$$B_{c2}(T) = \frac{\phi_0}{2\pi \xi_{ab}^2(0)} \cdot \left(1 - \left(\frac{T}{T_c}\right)^2\right) \tag{14}$$

is shown in Fig. 3(a) and free-fitting parameters were derived giving $T_c = 4.57 \pm 0.03$ K and $\xi_{ab}(0) = 17.3 \pm 0.1$ nm.

Substituting the derived coherence length, $\xi_{ab}(0) = 17.3$ nm, into $\kappa_c = \lambda_{ab}(0)/\xi_{ab}(0)$ and using Eq. 11 allows us to fit the $J_c$ data and derive thermodynamic parameters for the Ga film as follows: the transition temperature, $T_c = 4.40 \pm 0.12$ K, the specific heat jump at $T_c$, $\Delta C/C = 0.84 \pm 0.18$, the London penetration depth, $\lambda_{ab}(0) = 371.5 \pm 6$ nm, the superconducting energy gap, $\Delta(0) = 0.91 \pm 0.20$ meV, and GL parameter, $\kappa_c = 21.5$ (Fig. 3,b).

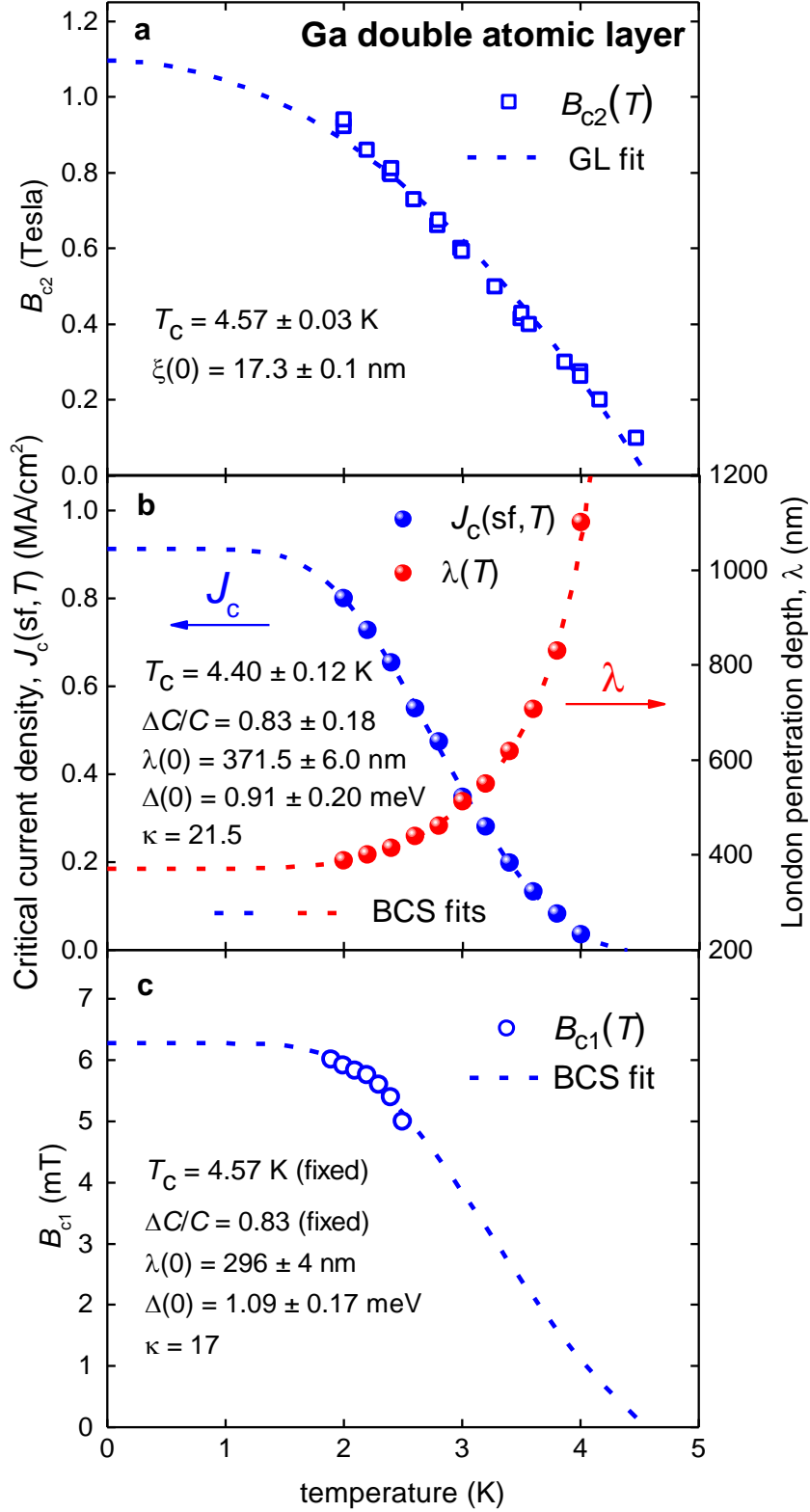

**Figure 3.** Experimental temperature dependence of (a) $B_{c2}(T)$, (b) $J_c(sf,T)$ and (c) $B_{c1}(T)$ data together with fits as described in the text for double-atomic-layer hexagonal Ga films ($2b$ = 0.552 nm). (a) Dashed curve is GL fit to Eq. 14. Derived parameters are $T_c = 4.57 \pm 0.03$ K and $\xi_{ab}(0) = 17.3 \pm 0.1$ nm, $R = 0.9881$. (b) Dashed curve is BCS fit using Eq. 11. Derived parameters are: $T_c = 4.40 \pm 0.12$ K, $\Delta C/C = 0.83 \pm 0.20$, $\lambda_{ab}(0) = 371.5 \pm 6.0$ nm, $\Delta(0) = 0.91 \pm 0.20$ meV, and $\kappa_c = 21.5$ ($\xi(0) = 17.3$ nm was fixed), $R = 0.9826$. (c) Dashed curve is BCS

fit to Eq. 15. Derived parameters are: $\lambda_{ab}(0) = 296 \pm 4$ nm, $\Delta(0) = 1.09 \pm 0.17$ meV, and $\kappa_c = 17$ ($T_c = 4.57$ K, $\Delta C/C = 0.83$, $\xi_{ab}(0) = 17.3$ nm were fixed), $R = 0.9983$.

Note that the derived superconducting energy gap is in excellent agreement with that reported from differential tunneling conductance spectra for the same film [15], which gave $\Delta(0) = 1.01 \pm 0.05$ meV.

We can also fit the $B_{c1}(T)$ data set from the same work [15] to the Ginzburg-Landau expression:

$$B_{c1}(T) = \frac{\phi_0}{4\pi \lambda_{ab}^2(T)} \cdot \left( \ln\left(\frac{\lambda_{ab}(0)}{\xi_{ab}(0)}\right) + 0.5 \right) \tag{15}$$

where $\lambda(T)$ is calculated again using Eq. (8) and we use $\xi_{ab}(0) = 17.3$ nm. As the raw $B_{c1}(T)$ data were limited to six values in the temperature interval of $T = 1.9 - 2.5$ K, we therefore fixed the transition temperature to the value $T_c = 4.57$ K (obtained from the $B_{c2}(T)$ fit), and the specific heat jump to the value $\Delta C/C = 0.84$ (obtained from $J_c(sf,T)$ fit). That left in this case just $\lambda(0)$ and $\Delta(0)$ as free fitting parameters.

The fit to the $B_{c1}(T)$ data is shown in Fig. 3(c) and, despite the limited data, the derived London penetration depth, $\lambda(0) = 296 \pm 4$ nm as well as the superconducting energy gap, $\Delta(0) = 1.09 \pm 0.17$ meV, are in very good agreement with the corresponding parameters obtained from the $J_c(sf,T)$ fit and differential tunneling conductance spectra technique, respectively.

5. **Single atomic layer FeSe**

The superconducting transition temperature of single-atomic-layer FeSe ($2b = 0.55$ nm) is $T_c = 23.5$ K [14] and it is remarkably higher than the transition temperature of bulk FeSe crystals with $T_c = 8$ K. Self-field critical currents were measured on a current bridge having width $2a = 1.45$ mm [14]. For $J_c(sf,T)$ analysis we used the GL parameter $\kappa = 72.3$ [44] found for bulk FeSe crystals.

A fit of the available $J_c(\text{sf},T)$ data to a single-band BCS model (Eqs. 10 and 11) is shown in Fig. 4,a. The fit is reasonably good (with $R = 0.9682$), and derived parameters match well the values obtained for bulk samples, especially the London penetration depth, $\lambda(0) = 335 \pm 1$ nm, which is in remarkable agreement with the bulk value, $\lambda(0) = 325$ nm [44].

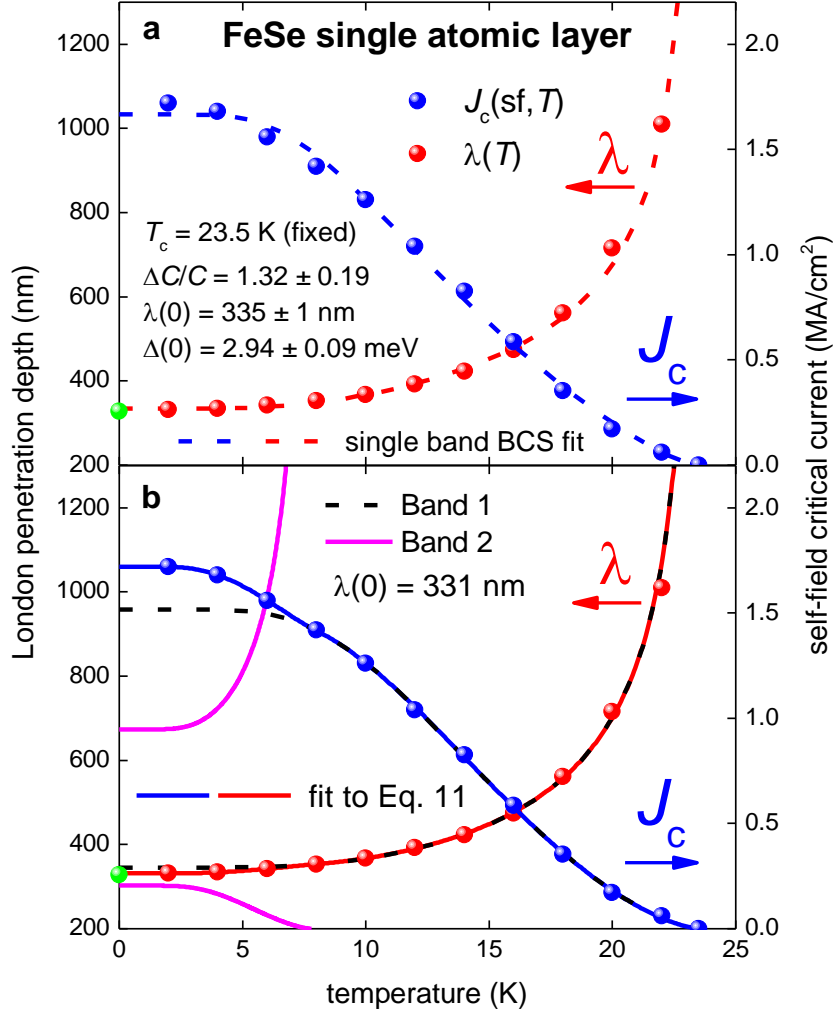

**Figure 4.** Experimental $J_c(\text{sf},T)$ data for a single-atomic-layer film of FeSe ($2a = 1.5$ mm, $2b = 0.55$ nm) fitted to (a) the single- and (b) the two-superconducting-band BCS model ($\kappa = 72.3$ [44]). The single green data point at $T = 0$ K is the independently-reported ground-state value of $\lambda_{ab}(0) = 325$ nm [44]. (a) The dashed curve is the BCS fit to Eqs. 10 and 11. Derived parameters are $\Delta C/C = 1.32 \pm 0.19$, $\lambda_{ab}(0) = 335 \pm 1$ nm, $\Delta(0) = 2.94 \pm 0.09$ meV ($T_c = 23.5$ K was fixed as an experimental data point), $R = 0.9682$. (b) Solid lines are the fit to the two-decoupled-bands model (Eq. 13). The ground-state composite London penetration depth is $\lambda_{ab}(0) = 331$ nm. The derived parameters for Band 1 are: $T_{c1} = 23.6 \pm 0.4$ K, $\Delta C_1/C_1 = 1.1 \pm 0.1$, $\lambda_1(0) = 346 \pm 2$ nm, $\Delta_1(0) = 3.45 \pm 0.13$ meV, and $2\Delta_1(0)/k_B T_{c1} = 3.4 \pm 0.2$; while the derived parameters for Band 2 ("bulk"-like band) are: $T_{c2} = 7.8 \pm 0.5$ K, $\lambda_2(0) = 674 \pm 28$ nm, $\Delta_2(0) = 1.4 \pm 0.3$ meV ($\Delta C_2/C_2 = 1.43$ was fixed to the BCS weak-coupling value), $2\Delta_2(0)/k_B T_{c2} = 4.2 \pm 1.0$, $R = 0.9926$.

However, a much better fit was obtained using the model of two decoupled bands ($R = 0.9926$). The fit is shown in Fig. 4(b). It is intriguing to find that the band with the smaller gap has more or less identical parameters to the bulk FeSe superconductor. For instance, it has a critical temperature $T_{c2} = 7.8 \pm 0.5$ K, remarkably close to the usual values of $T_c = 7.9$-$8.3$ K [13, 44] reported for bulk FeSe. Moreover, the derived superconducting energy gap, $\Delta(0) = 1.4 \pm 0.3$ meV, sits within the range of values 1.23 meV [45] to 2.2 meV [46] reported for bulk FeSe crystals. ARPES measurements on FeTe$_{0.6}$Se$_{0.4}$ [47] perhaps clarify this variation. The gap is anisotropic in the basal-plane Brillouin zone, modulating with four-fold rotational symmetry between 1.22 meV and 2.0 meV. Our deduced value is therefore very reasonable. Any anisotropy would simply modify the detailed $T$-dependence of $J_c$(sf) below $T_{c2}$ which, in our case, is insufficiently defined by just three low-$T$ data points.

The band with the larger gap with $T_{c1} = 23.6 \pm 0.4$ K and $\lambda_{ab,1}(0) = 346 \pm 2$ nm has parameters that one can expect for a weak-coupled BCS superconductor: $\Delta C_1/C_1 = 1.1 \pm 0.1$, and $\Delta_1(0) = 3.45 \pm 0.13$ meV, that converts to $2\Delta_1(0)/k_B T_{c1} = 3.40 \pm 0.15$, close to the weak-coupling BCS value. The derived total London penetration depth (which is the composite value originating from both bands) $\lambda_{ab}(0) = 331$ nm is even closer to the bulk value $\lambda(0) = 325$ nm [44], than from the single-band fit (Fig. 4,a). ARPES and STM measurements on single- or few-atomic layer FeSe do indeed reveal a second, larger gap (without revealing the coexisting smaller gap we identify here – it is only seen in thicker samples). In one study [48] this larger gap rises from 9 meV to 15 meV as the film thickness is reduced to one unit cell, much larger than the 3.45 meV we deduce. This difference is possibly attributable to the high gap value observed *in situ* in single-unit-cell films with estimated $T_c$ above 65 K compared with the much smaller value of zero-resistance $T_c$ around 24 K detected by *ex situ* transport measurements as reported here (though still enhanced over the bulk value of around 8 K). The nature of the substrate seems to play a role here [48]. Moreover if $T_c$ is reduced in our case by scattering and fluctuations this will also be reflected in an apparent reduced gap magnitude. The important point is that a second larger gap is observed in ARPES and STM studies and, perhaps more importantly, the decay length observed for the enhanced gap [48] is found to be of the order of the $c$-axis coherence length as we discuss in more detail later.

6. **Exfoliated 2H-TaS$_2$**

2H-TaS$_2$ is a layered superconductor with inter-plane distance of 0.60 nm [49]. Studies of the transition temperatures for exfoliated crystals of 2H-TaS$_2$ showed that there is a pronounced enhancement in $T_c$ from 0.5 to 2.2 K as the crystals are thinned down from thickness of $2b = 14.9$ nm to $2b = 3.5$ nm [17]. A literature search for the Ginsburg-Landau parameter $\kappa$ for 2H-TaS$_2$ reveals a quite large scatter, i.e., $\kappa = 9.5$ [50], 9.8 [51], 13.6 [52], 12.1 [52], 15.1 [52], 4.2 [53], with an average value of $\kappa = 10.7 \pm 3.9$. From this mean value we can estimate a range of expected $\lambda(0)$, based on the measured value of $B_{c2}(0) = 0.11$ T [17], which converts by Eq. 15 into an in-plane coherence length of $\xi_{ab}(0) = 54.7$ nm. Thus, the range of values for the London penetration depth is expected to be $\lambda_{ab}(0) = \kappa_c \cdot \xi_{ab}(0) = 585 \pm 213$ nm. For $J_c(\text{sf},T)$ fits we use the rounded value for the GL parameter $\kappa = 11$. Note that as $\kappa$ is under the logarithm in Eq. (2) then the accuracy of derived parameters from $J_c(\text{sf},T)$ will be little effected by the uncertainty in $\kappa$.

Fig. 5(a) shows a fit of the $J_c(\text{sf},T)$ data for a 3.5 nm thick ($2a = 1,000$ nm) 2H-TaS$_2$ crystal to the single-band BCS model (Eqs. 10 and 11).

The fit is very good (with $R = 0.9790$) and the derived energy gap, $\Delta(0) = 277 \pm 7$ µeV, is in remarkably good agreement with the reported value of $\Delta(0) = 280$ meV [54] found by scanning tunneling spectroscopy studies for 2H-TaS$_2$ single crystals. The derived London penetration depth $\lambda(0) = 733 \pm 2$ nm is in the expected range of $585 \pm 213$ nm.

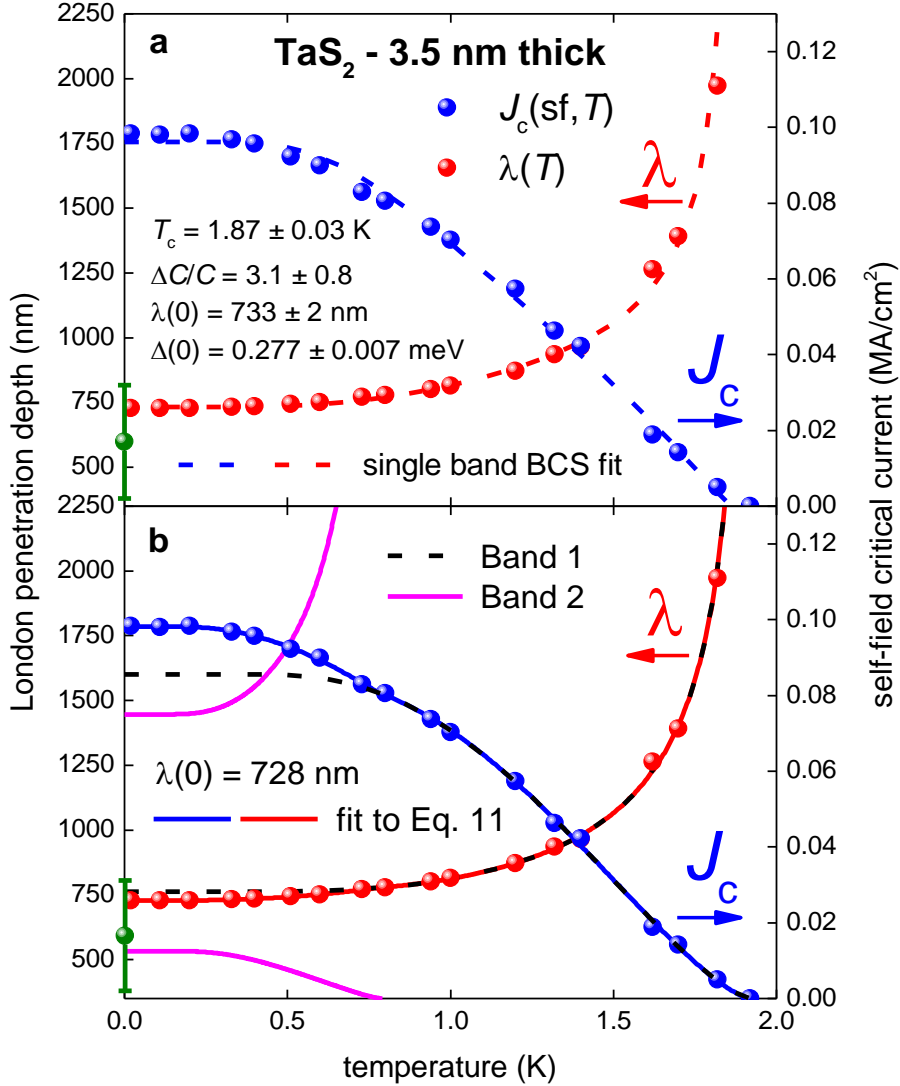

**Figure 5.** Experimental $J_c(\text{sf},T)$ data for a 2H-TaS$_2$ exfoliated crystal ($2a = 1,000$ nm, $2b = 3.5$ nm) fitted to (a) single-superconducting-band model (Eqs. 10 and 11), and (b) the two-decoupled-superconducting-bands model (Eq. 13). The Ginzburg-Landau parameter is $\kappa_c = 11$. The single green data point with error bar at $T = 0$ K is the value of $\lambda_{ab}(0)$ calculated from the experimentally-measured coherence length $\xi_{ab}(0) = 54.7$ nm and $\kappa_c = 10.7 \pm 3.9$, giving $\lambda_{ab}(0) = \kappa_c \cdot \xi_{ab}(0) = 585 \pm 213$ nm. (a) The dashed curve is the BCS fit to Eqs. 10 and 11. Derived parameters are $T_c = 1.87 \pm 0.03$ K, $\Delta C/C = 3.1 \pm 0.8$, $\lambda_{ab}(0) = 733 \pm 2$ nm, $\Delta(0) = 277 \pm 7$ μeV, $R = 0.9790$. (b) Solid lines are a fit to the two-decoupled-bands model with derived total ground-state London penetration depth, $\lambda_{ab}(0) = 728$ nm. Derived parameters for Band 1 are: $T_{c1} = 1.92 \pm 0.03$ K, $\Delta C_1/C_1 = 2.0 \pm 0.3$, $\lambda_1(0) = 762 \pm 6$ nm, $\Delta_1(0) = 0.35 \pm 0.02$ meV, and $2\Delta_1(0)/k_B T_{c1} = 4.23 \pm 0.11$; derived parameters for Band 2 ("bulk"-like band): $T_{c2} = 0.790 \pm 0.055$ K, $\lambda_2(0) = 1446 \pm 81$ nm, $\Delta_2(0) = 0.131 \pm 0.018$ meV ($\Delta C_2/C_2 = 2.0$ was fixed to the value derived for Band 1), and $2\Delta_2(0)/k_B T_{c2} = 3.8 \pm 0.5$, $R = 0.9930$.

However, it can be seen (Fig. 5,a) that at low temperatures the experimental $J_c$(sf,$T$) data behaves very similarly to that seen in the FeSe superconductor (Fig. 4,a), i.e., an additional rise in critical current occurs at $T \sim (0.3\text{-}0.4)T_c$. A fit of $J_c$(sf,$T$) to the two-decoupled-bands model is excellent and it is shown in Fig. 5,b. There are two important issues here. The first is that the fit to the two-coupled-bands model does not converge. The second is that the decoupled-bands fit can be made when all 8 parameters are free. However, because of the limited $J_c$(sf,$T$) data set the derived parameters have quite large uncertainties, especially $\Delta C_2/C_2$ for which the uncertainty is larger than the derived value. So, we reduced the number of free parameters by one, by assuming that $\Delta C_2/C_2$ is equal to the free-fitting parameter of $\Delta C_1/C_1$. We note that $\Delta C_2/C_2$ can be assumed to be equal to the weak-coupling BCS limit without any significant changes in values for other derived parameters. But we made an attempt to use a more flexible approach as 2H-TaS$_2$ is likely to be more strongly-coupled. We use the same approach for all 2H-TaS$_2$ fits herein.

As a result, for the two-decoupled-bands model the derived ground-state London penetration depth (which is the collective value arising from both bands) is $\lambda_{ab}(0) = 728$ nm, practically the same as the value derived for the single-band model (Fig. 4,a). Both bands, within uncertainty intervals, show a higher $2\Delta_i(0)/k_BT_{ci}$ ratio than the BCS weak-coupling limit of 3.53. The derived value of $\Delta C_1/C_1$ is also above the BCS weak-coupling limit of 1.43, which supports the overall result for the fit ($R = 0.9930$). And again, as for FeSe, we have the intriguing finding that the smaller band has a very similar transition temperature, $T_{c2} = 0.79 \pm 0.06$ K, to the bulk 2H-TaS$_2$ crystals.

To further explore this finding, we measure and fit $J_c$(sf,$T$) for as wide a range of sample thicknesses as experimentally possible, in this case from $2b = 4.2$ nm to $2b = 14.9$ nm. The results are presented in Fig. 6 and Table I. For most of these exfoliated crystals the penetration depth is within or close to the expected value of $\lambda_{ab}(0) = 585 \pm 213$ nm. The variation can be attributed to some level of influence from weak links in the form of partial delamination.

The 'bulk-like' band for all these 2H-TaS$_2$ crystals has a critical temperature within a relatively narrow interval of $T_c = 0.70 \pm 0.07$ K. The consistency of values for the energy gap for this band, $\Delta(0) = 137 \pm 24$ µeV, is also evident. The small apparent trend of increasing transition temperature for each band as the crystal thickness reduces may not be significant. The standard deviations in $T_c$ values shows that they are largely within error of each other. However, as noted, the larger energy gap in ultra-thin FeSe varies with thickness [48] and it would not be surprising if a similar effect were evident in TaS$_2$.

**Table I.** Derived parameters for 2H-TaS$_2$ ultrathin crystals.

| TaS$_2$ crystal thickness, $2b$ (nm) | 3.5 | 4.2 | 5.8 | 10.2 | 14.9 |
|---|---|---|---|---|---|
| $\lambda(0)$ (nm) | 728 | 395 | 856 | $1041 \pm 14$ | $866 \pm 8$ |
| $T_{c1}$ (K) | $1.92 \pm 0.03$ | $1.88 \pm 0.07$ | $1.47 \pm 0.03$ | | |
| $\Delta C_1/C_1$ | $2.0 \pm 0.3$ | $1.31 \pm 0.43$ | $2.1 \pm 0.4$ | | |
| $\lambda_1(0)$ (nm) | $762 \pm 6$ | $419 \pm 6$ | $893 \pm 25$ | | |
| $\Delta_1(0)$ (µeV) | $349 \pm 24$ | $345 \pm 54$ | $221 \pm 25$ | | |
| $2\Delta_1(0)/k_BT_{c1}$ | $4.2 \pm 0.1$ | $4.3 \pm 0.7$ | $3.49 \pm 0.45$ | | |
| $T_{c2}$ (K) | $0.79 \pm 0.06$ | $0.70 \pm 0.10$ | $0.58 \pm 0.16$ | 0.75 | $0.65 \pm 0.03$ |
| $\Delta C_2/C_2$ | 2.0 (fixed) | 1.3 (fixed) | 2.1 (fixed) | $2 \pm 1$ | $2.2 \pm 1.2$ |
| $\lambda_2(0)$ (nm) | $1446 \pm 81$ | $720 \pm 53$ | $1757 \pm 385$ | $1041 \pm 14$ | $866 \pm 8$ |
| $\Delta_2(0)$ (µeV) | $131 \pm 18$ | $138 \pm 75$ | $85 \pm 22$ | $157 \pm 46$ | $159 \pm 50$ |
| $2\Delta_2(0)/k_BT_{c2}$ | $3.8 \pm 0.5$ | $4.6 \pm 2.5$ | $3.4 \pm 0.9$ | $4.9 \pm 1.5$ | $5.7 \pm 1.8$ |

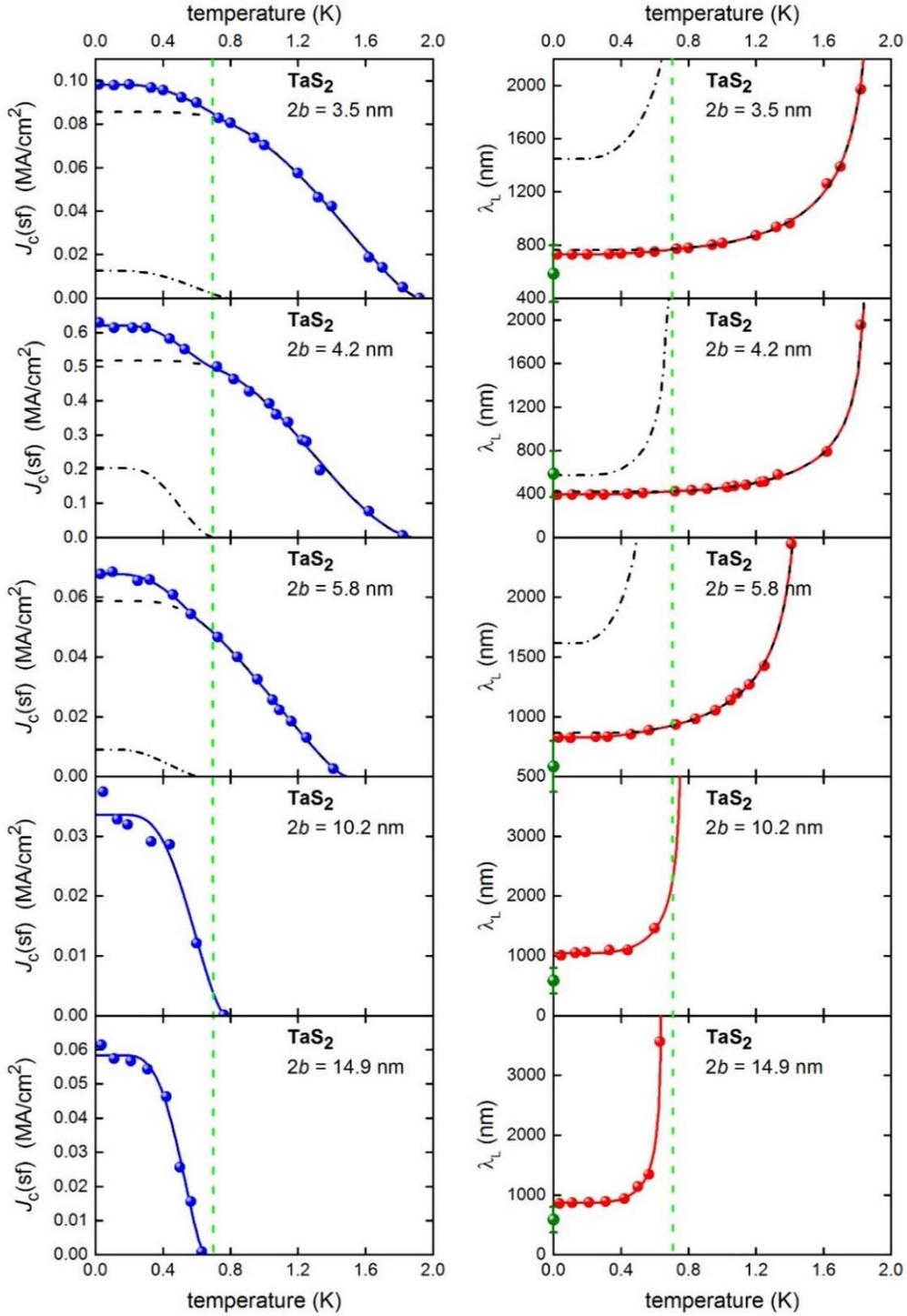

**Figure 6.** Experimental $J_c(\text{sf},T)$ for 2H-TaS$_2$ crystals (left column), the deduced London penetration depth, $\lambda(T)$, (right column), and the fits to the two-decoupled-bands model, and single-band-model (g-l). The 2H-TaS$_2$ crystal have cross-section dimensions, respectively, of $2a = 1{,}000$ nm and $2b = 3.5$ nm (a,b) ($R = 0.9930$), $2a = 450$ nm and $2b = 4.2$ nm (c,d) ($R = 0.9705$), $2a = 400$ nm and $2b = 5.8$ nm (e,f) ($R = 0.9816$), $2a = 1{,}370$ nm and $2b = 10.2$ nm (g,h) ($R = 0.5886$), and $2a = 2{,}000$ nm and $2b = 14.9$ nm (k,l) ($R = 0.7445$), as annotated. The contribution of Band 1 is shown by the dashed curves, while the contribution of Band 2 is shown by the dash-dot curves. Fitting parameters are listed in Table I.

### 7. Proposed criterion

Based on these results obtained for ultrathin FeSe and TaS$_2$ crystals we can ask the question: is there a common physical condition at which the new superconducting band appears with decreasing crystal thickness? Considering many possibilities for these very different superconductors, we found that the common circumstance is that the crystal thickness becomes smaller than the ground-state out-of-plane coherence length, $\xi_c(0)$. Consider the FeSe single-atomic-layer superconductor, noting that:

$$\xi_c(0) = \frac{\xi_{ab}(0)}{\gamma(0)} = \frac{\lambda_{ab}(0)}{\kappa_c\,\gamma(0)} \qquad (16)$$

where $\gamma(0)$ is the mass anisotropy (which for FeSe is 2.0 [55,56],) then, by using our derived $\lambda_{ab}(0) = 311$ nm, and $\kappa = 72.3$ [44], we find an out-of-plane coherence length of $\xi_c(0) = 2.15$ nm. This value is four times larger than the thickness of single-atomic-layer FeSe of $2b = 0.55$ nm. Based on this, we can expect that the second large-gap in FeSe will close when the film thickness exceeds four FeSe monoatomic layers (ML). This proposal is supported by photoemission studies performed by Tan and co-workers [57], where these authors found that at $T = 30$ K photoemission spectra are identical for films with thickness of 4 ML, 15 ML, and 35 ML. And there is a remarkable difference between these spectra and those for 1 ML, 2ML, and 3 ML films. Moreover, it has been reported that the enhancement in the larger energy gap for ultra-thin FeSe decays away over some 4-5 unit cells [48], again comparable to the magnitude of $\xi_c(0)$.

What might be the physical origins of this second gap? Firstly, there is clearly some electronic coupling between the superconducting films and their substrates. Using ARPES Lee *et al*. [58] have observed replica bands dispersing 100 meV below the originating bands in single-atomic-layer FeSe. These are attributed to bosonic modes, perhaps optical phonons, in the SrTiO$_3$ substrate that couple to electrons in FeSe, potentially opening and enhancing an energy gap. However, it is important to note that, in our studies, the original bulk gap does not appear to be affected and any model based on this coupling would have to recognise this fact. Alternatively the effect could be intrinsic, involving some kind of electronic renormalisation at the surface due, for example, to image Cooper pairs or coupling to surface plasmons.

Turning to the TaS$_2$ exfoliated crystals the mass anisotropy is $\gamma(0) = 6.7$ [59]. From the in-plane coherence length, $\xi_{ab}(0) = 54.7$ nm, the out-of-plane coherence length is found to be $\xi_c(0) = 8.2$ nm. This value just separates the two groups of samples: those with just a single "bulk" gap (samples thicker than $\xi_c(0)$) and those with (at least) two gaps (samples thinner than $\xi_c(0)$).

The data presented in Fig. 1 for the NbN film ($2b = 22.5$ nm) further supports our proposed idea, because for isotropic material $\xi_c(0) = \xi_{ab}(0)$, and in the case of NbN, $\xi_c(0) = \lambda(0)/\kappa = 194$ nm / 40 = 4.85 nm, which is much smaller than the film thickness of 22.5 nm. The inferred thermodynamic parameters are thus consistent with bulk values. Data presented in Fig. 2 for the MgB$_2$ film ($2b = 10$ nm) is less clear as this is a multi-band bulk superconductor. However, the absence of an additional gap and an enhanced $T_c$ above the bulk value does also support our proposal, because by taking $\gamma(0) = 2.5$, $\xi_c(0) = 85.7$ nm / (2.5×26) = 1.3 nm, which is also much smaller, than the film thickness of $2b = 10$ nm. And it is notable that for MgB$_2$ films the effect might be never be observable given the very small value of $\xi_c$.

Additionally, the experimental $J_c$(sf,$T$) data for double-atomic-layer hexagonal Ga ($2b = 0.552$ nm) does not cover a large enough temperature range to reveal the appearance of the expected second superconducting gap, and this system remains to be studied. However, we are proposing that, as $T_c$ of the double-atomic-layer Ga film is notably higher than for bulk

Ga, the same mechanism of the opening of an additional superconducting gap is likely to occur for both Ga double- and triple-atomic-layer [15,16] films.

**8. Hexagonal triple-atomic-layer gallium**

Hexagonal triple-atomic-layer Ga ($2b = 0.828$ nm) is a type-II superconductor [16] with resistive transition temperature ($R = 0$) of $T_c = 3.7$ K, which is lower than the resistive transition temperature of $T_c = 4.5$ K of double atomic layer of hexagonal Ga [15]. This reduction in $T_c$ with increase in film thickness concurs with our proposed idea (Section 7). Self-field critical currents were measured on a current bridge with width $2a = 2.0$ mm. To derive the coherence length we use the same approach as for the hexagonal double-atomic-layer Ga film, i.e. we applied a criterion of 50% normal resistance recovery to the $R(T, B_\perp)$ curves [16] to define the upper critical field, $B_{c2}(T)$, and fit $B_{c2}(T)$ data to the GL expression (Eq. 14), which are shown in Fig. 7,a.

The free-fitting value for the coherence length $\xi_{ab}(0) = 16.3 \pm 0.1$ nm is in good agreement with the value obtained for double-atomic-layer hexagonal Ga, $\xi_{ab}(0) = 17.3 \pm 0.1$ nm. Substituting the derived coherence length, $\xi_{ab}(0) = 17.3$ nm, in Eqs. 4 and 8 allows us to fit the $J_c$ data and derive thermodynamic parameters for the Ga film as follows: the transition temperature, $T_c = 3.95 \pm 0.03$ K, the specific heat jump at $T_c$, $\Delta C/C = 2.2 \pm 0.2$, the London penetration depth, $\lambda_{ab}(0) = 547 \pm 7$ nm, the superconducting energy gap, $\Delta(0) = 0.75 \pm 0.07$ meV, and GL parameter, $\kappa_c = 33.5$ (Fig. 7,b).

The lower energy gap, $\Delta(0)$, and larger penetration depth, $\lambda_{ab}(0)$, show the trend of weakening superconductivity in the triple-layer Ga film in comparison with the double-layer Ga film, consistent with the fall in transition temperature.

We should note that our proposed enhancement in $T_c$ in thin films due to the opening of a second superconducting gap (while the "bulk-like" gap remains unchanged) should be detectable by several other techniques which are sensitive to additional bands crossing the Fermi surface, such as scanning tunneling spectroscopy (STS) or ARPES, but also these distinct gaps should be evident in the temperature-dependence of the upper critical field.

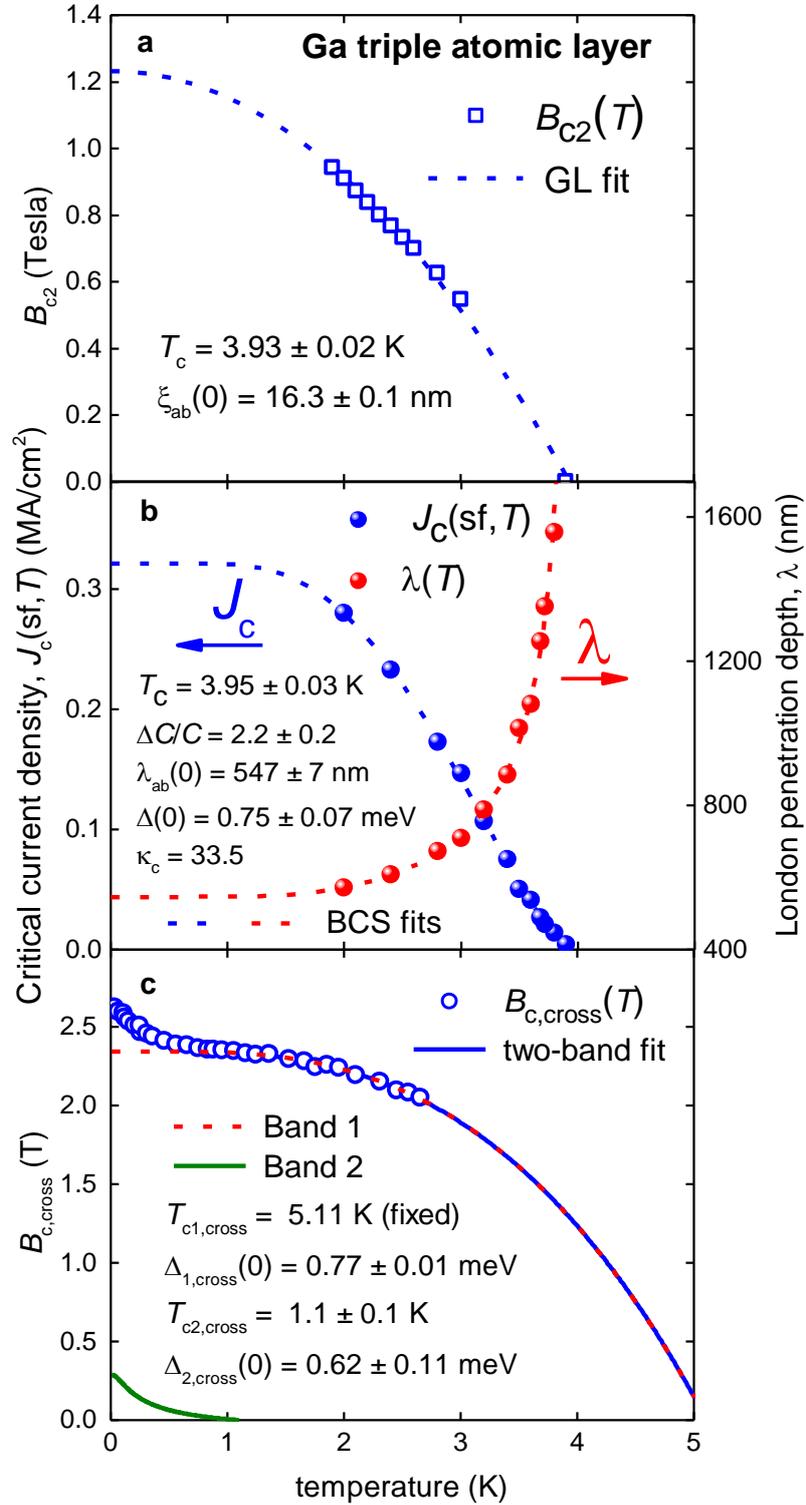

**Figure 7.** Experimental temperature dependence of (a) $B_{c2}(T)$, (b) $J_c(sf,T)$, and (c) $B_{c,cross}(T)$ with fits as described in the text for triple-atomic-layer hexagonal Ga films ($2b = 0.828$ nm). (a) The dashed curve is the GL fit to Eq. 14. Derived parameters are $T_c = 3.93 \pm 0.02$ K, $\xi_{ab}(0) = 16.3 \pm 0.1$ nm, and $R = 0.9970$. (b) The dashed curve is the BCS fit using Eqs. 10 and 11. Derived parameters are: $T_c = 3.95 \pm 0.03$ K, $\Delta C/C = 2.2 \pm 0.2$, $\lambda_{ab}(0) = 547 \pm 7$ nm, $\Delta(0) = 0.75 \pm 0.07$ meV, $\kappa_c = 33.5$, and $R = 0.9771$. (c) Fit to Eq. 17 with fixed $T_{c1,cross} = 5.11$ K. The annotation 'cross' refers to the field-dependent onset of resistance, so that $B_{cross}$ is a suitable proxy for $B_{c2}$. Derived parameters are: $\Delta_{1,cross}(0) = 0.77 \pm 0.01$ meV, $T_{c2,cross} = 1.1 \pm 0.1$ K, $\Delta_{2,cross}(0) = 0.62 \pm 0.11$ meV, and $R = 9996$.

Available $B_{c1}(T)$, and $B_{c2}(T)$ data for this system (Fig. 7) does not cover a sufficiently wide range of temperatures. However, there is field-dependent data for the on-set of the resistive superconducting transition in terms of the resistive crossover field, $B_{cross}(T)$, down to $T = 32$ mK [16]. Raw $B_{cross}(T)$ data show an upturn at about $T \sim 1$ K (Fig. 7,c). If we assume that $B_{cross}(T)$ can be treated similar to the upper critical field:

$$B_{cross}(T) = \frac{\phi_0}{2\pi \xi^2(T)} \tag{17}$$

and, if we adopt the simple assumption that the $T$-dependence of $\xi$ derives solely from that of $\Delta$, using $\xi = \hbar v_F/(\pi \Delta)$ then we may use Eq. 10 to calculate the $T$-dependence of $B_{cross}$. Thus, if $g(T)$ represents the $T$-dependence of $\xi^{-2}$ then the full equation for $B_{c2}(T)$ for a two-band superconductor is of the form:

$$B_{cross}(T) = \frac{\phi_0}{2\pi \xi_1^2(0)} \cdot g_1^2(T) + \frac{\phi_0}{2\pi \xi_2^2(0)} \cdot g_2^2(T) \tag{18}$$

where indices 1 and 2 denote Band 1 and Band 2, respectively.

A fit of $B_{cross}(T)$ to Eq. (18), where $T_{c1}$ is set to the experimental value of 5.11 K, is shown in Fig. 7,c. There is clear evidence that Band 2 has a transition temperature close to the bulk transition temperature of $T_c = 1.1$ K. And this inference will remain independent of the detailed model used to characterize the $T$-dependence of $B_{cross}$.

To lend more support for our proposal that a new superconducting gap appears when the crystal thickness become less than the out-of-plane coherence length, $\xi_c(0)$, we have searched the literature for $J_c$(sf,$T$) data for other very thin films. We report below analyses of datasets we could find to date.

## 9. Nb thin films

Rusanov et. al. [60] have reported $J_c$(sf,$T$) for very thin films of pure Nb. Fig. 8 shows raw $J_c$(sf,$T$) data along with our fit for Nb with film thickness of $2b = 20$ nm (we used $\kappa = 1.0$ [61]). The derived value of $\lambda(0) = 47.5$ nm combined with $\kappa = 1.0$ gives us $\xi(0) = 47.5$ nm, which is larger than the film thickness of $2b = 20$ nm.

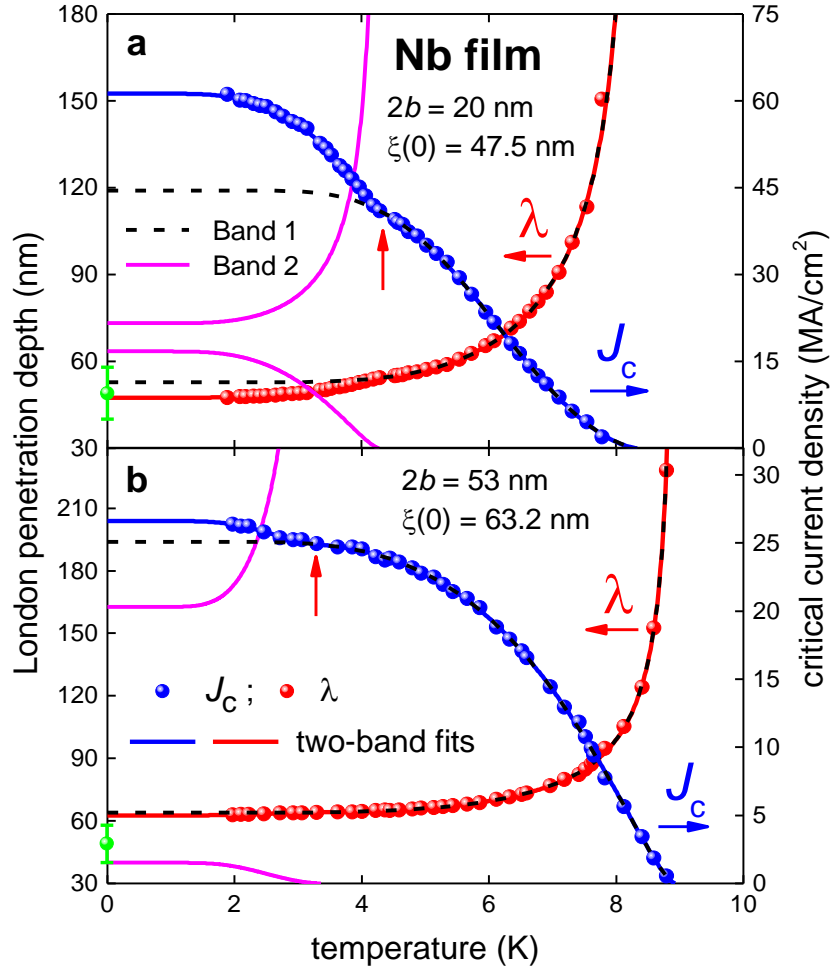

**Figure 8.** Experimental $J_c$(sf,$T$) data for thin films of Nb (right axis, blue) together with derived values of $\lambda(T)$ (left axis, red). The solid curves are fits using the two-decoupled-bands BCS-like model. The single green data point at $T = 0$ K is the independently-reported ground-state value of $\lambda(0) = 47$ nm [61]. (a) Thickness $2b = 20$ nm. Total London penetration depth $\lambda(0) = 47.5$ nm. Derived parameters for Band 1 are: $T_{c1} = 8.34 \pm 0.05$ K, $\Delta_1(0) = 2.45 \pm 0.26$ meV, $\Delta C_1/C_1 = 1.43 \pm 0.09$, $\lambda_1(0) = 52.8 \pm 0.2$ nm. Derived parameters for Band 2 are: $T_{c2} =$

4.28 ± 0.04 K, $\Delta_2(0)$ = 0.97 ± 0.06 meV, $\Delta C_2/C_2$ = 3.1 ± 0.3, $\lambda_2(0)$ = 73.2 ± 0.9 nm, $R$ = 0.9994. (b) – Thickness $2b$ = 53 nm. Total London penetration depth $\lambda(0)$ = 63.2 nm. Derived parameters for Band 1 are: $T_{c1}$ = 8.94 ± 0.02 K, $\Delta_1(0)$ = 2.12 ± 0.03 meV, $\Delta C_1/C_1$ = 3.7 ± 0.1, $\lambda_1(0)$ = 63.77 ± 0.06 nm. Derived parameters for Band 2 are: $T_{c2}$ = 3.5 ± 0.1 K, $\Delta_2(0)$ = 0.44 ± 0.29 meV, $\Delta C_2/C_2$ = 1.43 (fixed), $\lambda_2(0)$ = 135 ± 37 nm, $R$ = 0.9996.

The consequent appearance of a second gap is evident from the fits. For a thicker film, $2b$ = 53 nm (Fig. 8,b), this second band exhibits remarkably suppressed superconducting parameters including the transition temperature, and the energy gap. This observation is again well aligned with our general thesis that this band should disappear when the film thickness exceeds the coherence length.

### 10. Interface superconductors

We should mention that the interface superconductivity [62-65] which is found within interfaces of some oxides, might also originate from the same effect we infer here for ultra-thin superconducting films, namely the opening of a new superconducting gap because, similarly, the parent compounds (either side of the interface) generally have much lower $T_c$ values. If so, inasmuch as the interface film thickness becomes less than the out-of-plane coherence length, a new superconducting gap opens with much higher $T_c$ than the parent compounds.

To lend more support for this idea in Fig. 9 we show the critical temperature for a $(CaCuO_2)_n/(SrTiO_3)_{m=2}$ superlattice as a function of $CaCuO_2$ unit cell number, $n$, reported by Di Castro and co-workers [64]. The $CaCuO_2$ lattice constant is 0.384 nm [64], which means that when the $CaCuO_2$ sample thickness becomes less than 5 unit cells ($2b$ < 1.92 nm), $T_c$ is found to be enhanced. We note, that the $T_c$ drop at smaller $n$ has a different origin (more likely, a consequence of severe reduction of doping state).

There is no experimental data for the coherence length in $(CaCuO_2)_n/(SrTiO_3)_{m=2}$ superlattices, however, reported estimated values for the in-plane coherence length, $\xi_{ab}(0) = 2.5$-$3.5$ nm, and for the mass anisotropy, $\gamma = 4.5$-$7.5$, for a comparable $(CaCuO_2)_n/(SrTiO_3)_{m=3}$ superlattice [65], give grounds to expect that the enhancement in transition temperature in $(CaCuO_2)_n/(SrTiO_3)_{m=2}$ superlattices is likely to be associated with our proposed idea.

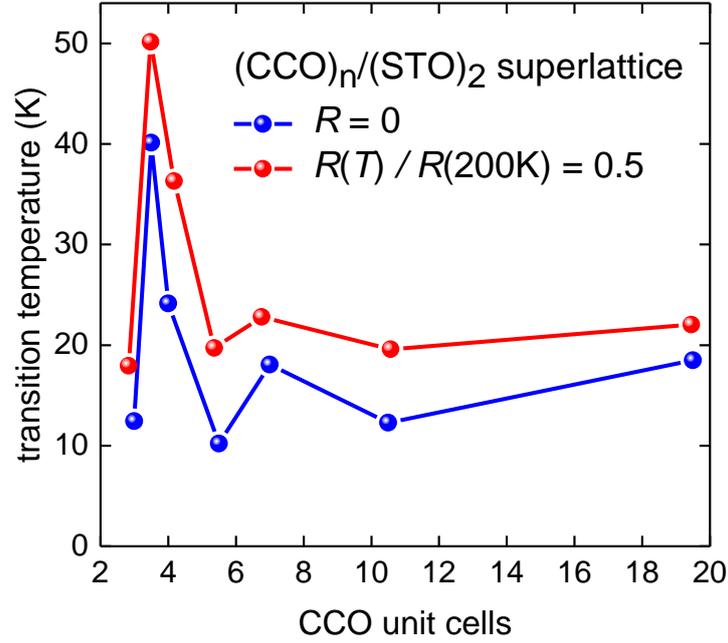

**Figure 9.** Transition temperature as a function of the number, $n$, of unit cells of the $CaCuO_2$ in $(CaCuO_2)_n/(SrTiO_3)_{m=2}$ superlattice, adopted from [65].

### 11. Possible systems for further studies

We consider now other 2D and atomically-thin systems in which the effect of a new gap opening might be easily observed. We have only chosen systems which are currently under active research and development. This does not mean that other systems beyond those listed below have less interest.

**13.1. 2H-MoS$_2$**

The Ye and Iwasa group discovered [66] that 2H-MoS$_2$, which is a bulk insulator, becomes a superconductor with highest transition temperature of about $T_c$ = 11 K, when it is thinned to several nanometers and then doped by the ionic-liquid gating (ILG) technique. More recently [67], these authors showed that ILG is a universal tool to induce superconductivity in many other transition metal dichalcogenides. Superconductivity has also been induced through proximity effect in single and few-layer MoS$_2$ flakes [68].

At present, only Costanzo *et al* [69] in their Figure 3 reported $I_c$(sf,$T$) and $B_{c2}(T)$ data for ILG transition metal dichalcogenides, namely bilayer 2H-MoS$_2$ ($2a$ = 20 μm, $2b$ = 1.23 nm) ion-gated at $V_{gate}$ = 2.2 V. Raw data and single-band fits for bilayer 2H-MoS$_2$ are shown in Fig. 10.

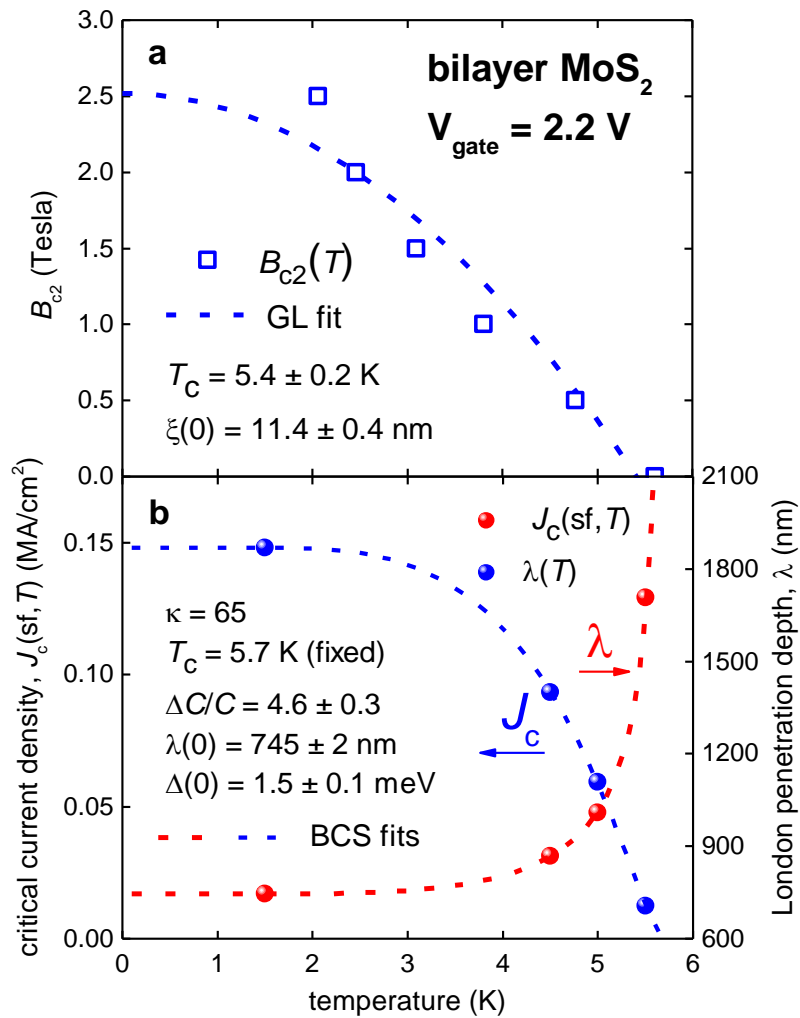

**Figure 10.** Experimental temperature dependence of (a) $B_{c2}(T)$ and (b) $J_c(sf,T)$ data together with fits as described in the text for a bilayer ILG MoS$_2$ film ($2b = 1.23$ nm) at $V_{gate} = 2.2$ V. (a) The dashed curve is the GL fit to Eq. 14. Derived parameters are $T_c = 5.4 \pm 0.2$ K and $\xi_{ab}(0) = 11.4 \pm 0.4$ nm. (b) Dashed curve is BCS fit using Eqs. 10 and 11. Derived parameters are: $\Delta C/C = 4.6 \pm 0.3$, $\lambda_{ab}(0) = 745 \pm 2$ nm, $\Delta(0) = 1.5 \pm 0.1$ meV, and $\kappa_c = 65$ (the values $\xi(0) = 11.4$ nm and $T_c = 5.7$ K were fixed).

The GL parameter was established to be $\kappa = 65$. The derived $\Delta C/C = 4.6 \pm 0.3$ and $\Delta(0) = 1.5 \pm 0.1$ meV, with $2\Delta(0)/k_B T_c = 6.1 \pm 0.5$ indicate that this superconductor is an extremely strong electron-phonon coupled superconductor. By looking at the raw $B_{c2}(T)$ data (Fig. 10(a)) we note that there is an indication that a new gap possibly opens at $T \sim 2.5 - 3.0$ K. More raw $I_c(sf,T)$ and $B_{c2}(T)$ data are required to make a more satisfactory analysis.

### 13.2. α-Mo$_2$C

Transition metal carbides form another class of 2D superconductors in which the effect of a new superconducting gap opening might be observed. Recently Xu *et al* [70] reported a reliable technology for manufacturing high-quality atomically-thin Mo$_2$C single crystals. From several $B_{c2}(T)$ and $I_c(st,T)$ datasets for Mo$_2$C films of different thicknesses and widths reported by Xu *et al* [70], we show in Fig. 11 processed data for a single crystal with $2a = 9.5$ μm and $2b = 7.5$ nm ($B_{c2}(T)$ data were presented in Fig. 3(c) of [70], and $I_c(st,T)$ data are from Fig. 4(a) of [70]).

The GL parameter was established to be $\kappa_c = 23$. The derived $\Delta C/C = 3.5 \pm 0.3$ and $\Delta(0) = 0.61 \pm 0.08$ with $2\Delta(0)/k_B T_c = 4.8 \pm 0.6$ indicate that this superconductor is a strong - coupled superconductor. However, measurements need to be done below 2 K to ascertain whether a second gap opens.

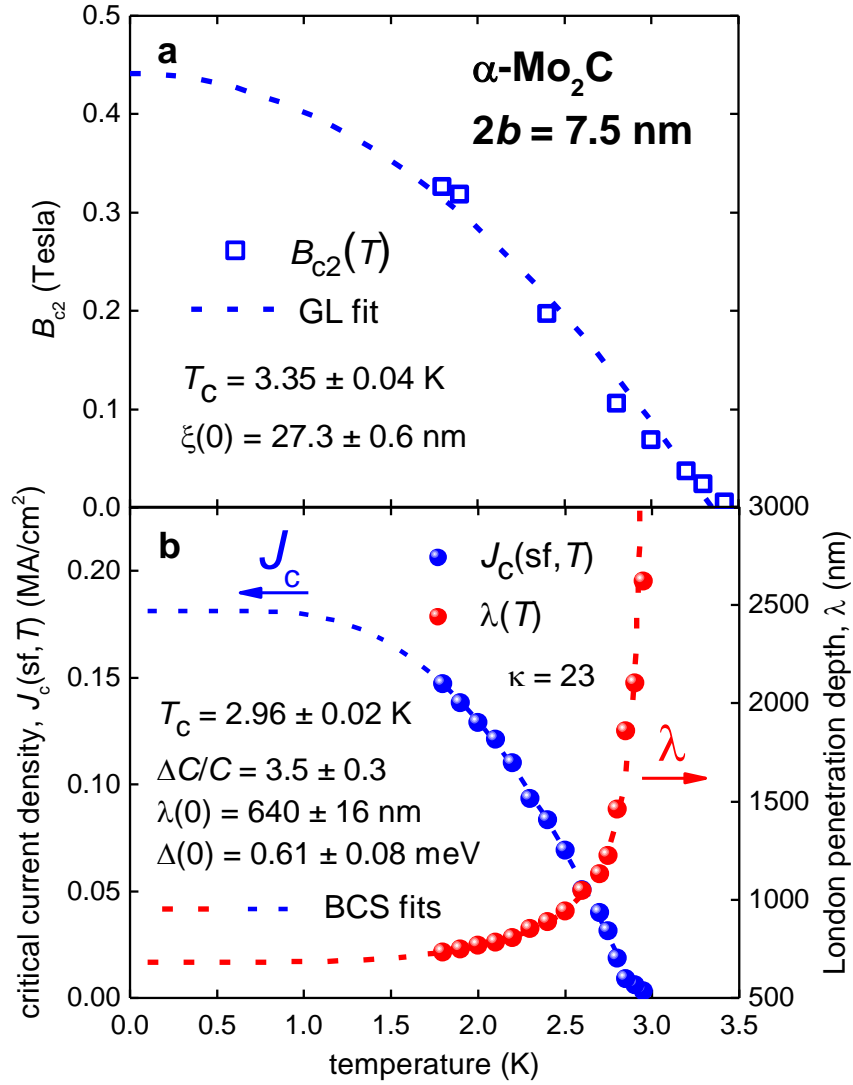

**Figure 11.** Experimental temperature dependence of (a) $B_{c2}(T)$ and (b) $J_c(sf,T)$ data together with fits as described in the text for Mo$_2$C film ($2b = 7.5$ nm). (a) The dashed curve is GL fit to Eq. 14. Derived parameters are $T_c = 3.35 \pm 0.04$ K and $\xi_{ab}(0) = 27.3 \pm 0.6$ nm; $R = 0.9780$. (b) The dashed curve is a BCS-like fit using Eqs. 10 and 11. Derived parameters are: $T_c = 2.96 \pm 0.02$ K, $\Delta C/C = 3.5 \pm 0.3$, $\lambda_{ab}(0) = 640 \pm 16$ nm, $\Delta(0) = 0.61 \pm 0.08$ meV, $2\Delta(0)/k_B T_c = 4.8 \pm 0.6$, $\kappa_c = 23$ (based on $\xi_{ab}(0) = 27.3 \pm 0.6$ nm); $R = 0.9776$.

### 13.3. NbSe$_2$

Niobium diselenide is another 2D superconductor in which the effect of new superconducting band opening might be observed. Recently, several groups [71-73] were successful in manufacturing high-quality atomically-thin crystals. Yoshida *et al* [73] reported that the transition temperature of atomically-thin crystals of NbSe$_2$ can be tuned by the ILG technique. From several available $B_{c2}(T)$ and $I_c(st,T)$ datasets for single, bilayer, trilayer, 4-layer, and 8-layer NbSe$_2$ crystals [71,72], we show in Fig. 12 $B_{c2}(T)$ and $J_c(st,T)$ data and fits for bilayer Sample #103 reported by Xi *et al* [72] (from their Figs. S4(b) and S7(a)). The derived parameters are in the expected range for a moderately strong-coupling superconductor, which NbSe$_2$ is. The derived London penetration depth, $\lambda_{ab}(0) = 250 \pm 60$ nm, is within its uncertainty of the reported value $\lambda_{ab}(0) = 200$ nm [74]. It is clear that low temperature $I_c(st,T)$ data are essential to reduce the uncertainty of the derived parameters for our model and to see if there is evidence of a second low-temperature gap opening, as is suggested by the data.

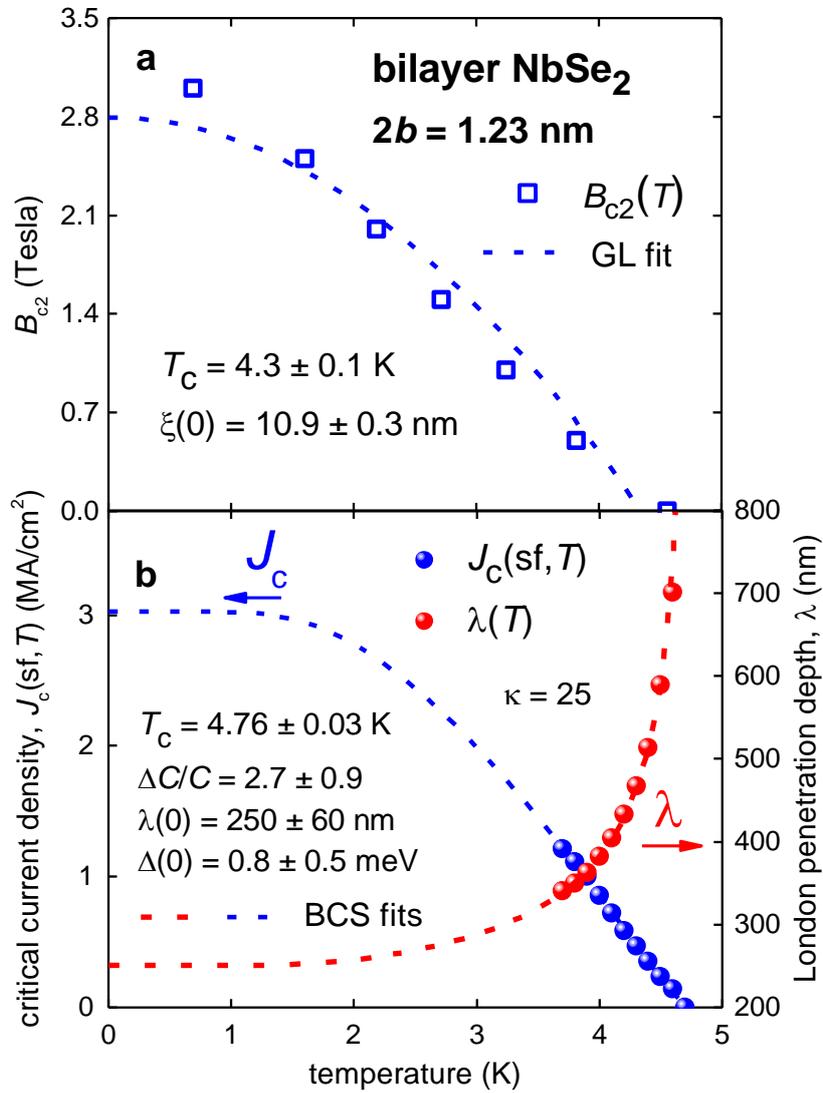

**Figure 12.** Experimental temperature dependence of (a) $B_{c2}(T)$ and (b) $J_c(sf,T)$ data together with fits as described in the text for bilayer NbSe$_2$ film ($2b = 1.23$ nm). (a) Dashed curve is the GL fit to Eq. 14. Derived parameters are $T_c = 4.3 \pm 0.1$ K and $\xi_{ab}(0) = 10.9 \pm 0.3$ nm; $R = 0.9597$. (b) The dashed curve is a BCS-like fit using Eqs. 10 and 11. Derived parameters are: $T_c = 4.73 \pm 0.03$ K, $\Delta C/C = 2.7 \pm 0.9$, $\lambda_{ab}(0) = 250 \pm 60$ nm, $\Delta(0) = 0.8 \pm 0.5$ meV, $2\Delta(0)/k_B T_c = 3.9 \pm 2.5$, $\kappa_c \approx 25$ (based on $\xi_{ab}(0) = 10.9 \pm 0.3$ nm); $R = 0.9741$.

### 13.4. Cuprates

All materials considered to this point were type-II *s*-wave superconductors. High-temperature superconducting cuprates form the widest class of quasi-2D superconductors which are, however, type-II *d*-wave superconductors. As there is a vast literature on cuprate superconductors we defer any discussion on these with the exception of one very recent report.

Fete *et al* [32] deduced $\lambda(T = 4.2\text{K})$ for ILG four-unit-cell-thick YBa$_2$Cu$_3$O$_7$ films by the same approach (Eq. (2)) and showed that the transition temperatures, $T_c$, and deduced $\lambda(T = 4.2\text{K})$ for these films follow the universal Uemura relation [33]. This is another promising demonstration that the ILG technique could be useful for revealing the effect of additional gap opening in superconductors a few atomic layers thick.

### 13.5. ZrNCl-EDLT

Saito *et al* [75] recently showed that superconductivity can be induced in ultra-thin films of the archetypical band insulator ZrNCl with transition temperature up to $T_c = 15$ K by ILG. The status of the topic was recently reviewed [76]. Although critical current data for ZrNCl is unavailable, this compound is another potential candidate for observing and studying the additional-gap effect we have proposed herein.

## 12. Conclusions

Here we have analyzed self-field critical currents for atomically-thin Ga, TaS$_2$ and FeSe superconductors and deduced their absolute values of the London penetration depth, the superconducting energy gap, and the relative jump in specific heat at $T_c$. It has been observed in all of these systems that $T_c$ is elevated relative to the bulk values and, in TaS$_2$ and FeSe, the enhancement in both cases has been previously attributed to increased electron-phonon interaction. Our central finding is that this enhancement in $T_c$ observed for these ultrathin materials arises from the opening of a second, larger superconducting gap, while keeping essentially unchanged the smaller "bulk" superconducting gap. The fact that this smaller gap remains unchanged is strong evidence that the electron-phonon interaction itself remains unchanged and that a new band moves up to cross the Fermi surface or a preexisting ungapped band at the Fermi surface becomes gapped. As such, the effect seems to be neither associated with the presence of a van Hove singularity [11] nor the effect of fluctuations. The effect for Ga double-atomic-layer films should be experimentally explored to lower temperatures than that currently available $T = 2$ K, as we expect that the "bulk"-like band remains with $T_c = 1.1$ K with a commensurate smaller gap in addition to the larger gap and $T_c$ value we deduce here. Searching the literature we find that a very thin niobium film [60] analyzed in the same manner also shows the presence of this second superconducting gap. Based on the available data, we conclude that for type-II superconductors the common physical condition at which this new superconducting band appears is that the film thickness falls below the ground-state out-of-plane coherence length for the material. A similar mechanism may also come into play for the appearance of interface superconductivity.

**Acknowledgement**

Authors thanks Dr. R. Gorbachev (National Graphene Institute, University of Manchester, U.K), Dr. X. Xi (Nanjing University, People's Republic of China), Prof. K. F. Mak (The Pennsylvania State University, USA), and Prof. W. Ren (Institute of Metal Research, Chinese Academy of Sciences, People's Republic of China) for providing raw experimental critical current datasets and supporting information for 2D superconductors analyzed in this work.

JLT thanks the Marsden Fund of New Zealand for financial support. EFT thanks Victoria University of Wellington for the support under Grant No. URF-8-1620-209864-3580 and PBRF Research Support Grant No. 215637. JOI acknowledges support by the Dutch organization for Fundamental Research on Matter (FOM), the Ministry of Education, Culture, and Science (OCW), and the Netherlands Organization for Scientific Research (NWO). JW, YS and YX thank the National Basic Research Program of China (Grants No. 2013CB934600), the Open Research Fund Program of the State Key Laboratory of Low-Dimensional Quantum Physics, Tsinghua University under Grant No. KF201501, and the Open Project Program of the Pulsed High Magnetic Field Facility (Grant No. PHMFF2015002), Huazhong University of Science and Technology.

**References**

1. Haviland D B, Liu Y and Goldman A. M. 1989 *Phys. Rev. Lett.* **62** 2180-2183
2. Yazdani A. and Kapitulnik A 1995 *Phys. Rev. Lett.* **74** 3037-3040
3. Guo Y *et al* 2004 *Science* **306** 1915–1917
4. Hermele M, Refael G, Fisher M P A and Goldbart P M 2005 *Nat. Phys.* **1** 117–121
5. Qin S, Kim J, Niu Q and Shih C K 2009 *Science* **324** 1314–1317
6. Jaeger H M, Haviland D B, Orr B G, and Goldman A M 1989 *Phys. Rev. B* **40** 182-196

7. Zaikin A D, Golubev D S, van Otterlo A, Zimányi G T 1997 *Phys. Rev. Lett.* **78** 1552-1555

8. Dubi Y, Meir Y, and Avishai Y 2007 *Nature* **449** 876-880

9. Refael G, Demler E, and Oreg Y 2009 *Phys. Rev. B* **79** 094524

10. Saito Y, Nojima T, and Iwasa Y 2016 *Nature Reviews Materials* **2** 16094

11. Hirsch J E and Scalapino D J 1986 *Phys. Rev. Lett.* **56** 2732-2735

12. Storey J G, Williams G V M and Tallon J L 2007 *Phys. Rev. B* **76** 174522

13. Wang Q-Y, Li Z, Zhang, W-H, Zhang Z-C, Zhang J-S, Li W, Ding H, Ou Y-B, Deng P, Chang K, Wen J, Song C-L, He K, Jia J-F, Ji S-H, Wang Y-Y, Wang L-L, Chen X, Ma X-C, Xue Q-K 2012 *Chin. Phys. Lett.* **29** 037402

14. Zhang W-H, Sun Y, Zhang J-S, Li F-S, Guo M-H, Zhao Y-F, Zhang H M, Peng J-P, Xing Y, Wang H-C, Fujita T, Hirata A, Li Z, Ding H, Tang C-J, Wang M, Wang Q-Y, He K, Ji S H, Chen X, Wang J-F, Xia Z-C, Li L, Wang Y-Y, Wang J, Wang L-L, Chen M-W, Xue Q-K, Ma X-C 2014 *Chin. Phys. Lett.* **31** 017401

15. Zhang H-M, Sun Y, Li W, Peng J-P, Song C-L, Xing Y, Zhang Q, Guan J, Li Z, Zhao Y, Ji S, Wang L, He K, Chen X, Gu L, Ling L, Tian M, Li L, Xie X C, Liu J, Yang H, Xue Q-K, Wang J, Ma X 2015 *Phys. Rev. Lett.* **114** 107003

16. Xing Y, Zhang H-M, Fu H-L, Liu H, Sun Y, Peng J-P, Wang F, Lin X, Ma X-C, Xue Q-K, Wang J, Xie X C 2015 *Science* **350** 542-545

17. Navarro-Moratalla E, Island J O, Manãs-Valero S, Pinilla-Cienfuegos E, Castellanos-Gomez A, Quereda J, Rubio-Bollinger G, Chirolli L, Silva-Guillén J A, Agraït N, Steele G A, Guinea F, Van Der Zant H S J, Coronado E 2016 *Nat. Comms.* **7** 11043

18. Wang Z, Liu C, Liu Y and Wang J 2017 *J. Phys.: Condens. Matter* **29** 153001

19. Talantsev E F and Tallon J L 2015 *Nat. Comms.* **6** 7820

20. Crump W P, Talantsev E F 2017 free-upload software:

    https://github.com/WayneCrump/BCS-theory-critical-current-fit

21. Poole P P, Farach H A, Creswick R J, Prozorov R *Superconductivity*, 2-nd Edition, London, UK, 2007, Chapter 6.

22. Shadowitz A 1981 *Phys. Rev. B* **24**, 2841-2843

23. Brandt E H, Indenbom M 1993 *Phys. Rev. B* **48**, 12893-12906

24. Clem J R, Bumble B, Raider S I, Gallagher W J, and Shih Y C 1987 *Phys. Rev. B* **35** 6637-6642

25. Poole P P, Farach H A, Creswick R J, Prozorov R *Superconductivity*, 2-nd Edition, London, UK, 2007, Chapter 12.

26. Poole P P, Farach H A, Creswick R J, Prozorov R *Superconductivity*, 2-nd Edition, London, UK, 2007, Chapter 14.

27. Bardeen J, Cooper L N, Schrieffer J R 1957 *Phys. Rev.* **108** 1175-1204

28. Mühlschlegel B 1959 Zeitschrift für Physik **155** 313–327

29. Won H and Maki K 1994 *Phys. Rev. B* **49** 1397-1402

30. Komiyama B, Wang Z, and Tonouchi M 1996 *Appl. Phys. Lett.* **68** 562-564

31. Kihlstrom K E, Simon R W, and Wolf S A 1986 *Phys. Rev. B* **32** 1843(R)- 1845(R)

32. Fête A, Rossi L, Augieri A, and Senatore C 2016 *Appl. Phys Lett.* **109** 192601.

33. Uemura Y J et al. 1989 *Phys. Rev. Lett.* **62** 2317-2320

34. Drozdov A P, Eremets M I, Troyan I A, Ksenofontov V, and Shylin S I 2015 *Nature* **525** 73-76

35. Talantsev E F, Crump W P, Storey J G, Tallon J L 2017 *Annalen der Physik (Berlin)* **529**, 201600390

36. Gross F, Chandrasekhar B S, Einzel D, Andres K, Hirschfeld P J, Ott H R, Beuers J, Fisk Z, Smith J L 1986 *Zeitschrift für Physik B Condensed Matter* **64** 175-188

37. Talantsev E F, Crump W P, and Tallon J L 2016 *arXiv* 1609.03670

38. Geibel C, Rietschel H, Junod A, Pelizzone M, and Muller J 1985 *J. Phys. F: Met. Phys.* **15**, 405-416

39. Bouquet F, Wang Y, Fisher R A, Hinks D G, Jorgensen J D, Junod A and Phillips N E 2001 *Europhys. Lett.* **56** 856-862

40. Luan L *et al* 2011 *Phys. Rev. Lett.* **106**, 067001

41. Zhang C, Wang Y, Wang D, Zhang Y, Feng Q-R, and Gan Z-Z 2013 *IEEE Trans. Appl. Supercond.* **23** 7500204

42. Panagopoulos C, Rainford B D, Xiang T, Scott C A, Kambara M, and Inoue I H 2001 *Phys. Rev. B* **64** 094514

43. Talantsev E, Strickland N; Hoefakker P, Xia J, Long N 2008 *Current Applied Physics* **8** 388-390

44. Lei H, Hu R and Petrovic C 2011 *Phys. Rev. B* **84** 014520

45. Khasanov R, Conder K, Pomjakushina E, Amato A, Baines C, Bukowski Z, Karpinski J, Katrych S, Klauss H-H, Luetkens H, Shengelaya A, and Zhigadlo N D 2008 *Phys. Rev. B* **78** 220510(R)

46. Song C L, Wang Y-L, Jiang Y-P, Li Z, Wang L, He K, Chen X, Ma X-C, and Xue Q-K 2011 *Phys. Rev. B* **84** 020503

47. Okazaki K, *et al*. 2012 *Phys. Rev. Lett.* **109** 237011

48. Zhang W H, Liu X, Wen C H P, Peng R, Tan S Y, Xie B P, Zhang T and Feng D L 2016 *Nano Lett.* **16** 1969-1973

49. Wagner K E, Morosan E, Hor Y S, Tao J, Zhu Y, Sanders T, McQueen T M, Zandbergen H W, Williams A J, West D V, and Cava R J 2008 *Phys. Rev. B* **78** 104520

50. Kashihara Y, Nishida A, and Yoshioka H 1979 *J. Phys. Soc. Jpn.* **46** 1112-1118

51. Gygax S 1979 *J. Low Temp. Phys.* **36** 109-119

52. Schwall R E, Stewart G R, and Geballe T H 1976 *J. Low Temp. Phys.* **22** 557–567

53. Garoche P, Manuel P, and Veyssie J J 1978 *J. Low Temp. Phys.* **30** 323–336

54. Guillamón I, Suderow H, Rodrigo J G, Vieira S, Rodière P, Cario L, Navarro-Moratalla E, Martín-Gastaldo C, Coronado E 2011 *New Journal of Physics* **13** 103020

55. Vedeneev S I, Piot B A, Maude D K, and Sadakov A V 2013 *Phys. Rev. B* **87** 134512

56. Abdel-Hafiez M, Zhang Y-Y, Cao Z-Y, Duan C-G, Karapetrov G, Pudalov V M, Vlasenko V A, Sadakov A V, Knyazev D A, Romanova T A, Chareev D A, Volkova O S, Vasiliev A N, and Chen X-J 2015 *Phys. Rev. B* **91** 165109

57. Tan S, Zhang Y, Xia M, Ye Z, Chen F, Xie X, Peng R, Xu D, Fan Q, Xu H, Jiang J, Zhang T, Lai X, Xiang T, Hu J, Xie B and Feng D 2013 *Nature Materials* **12**, 634–640

58. Lee J J, Schmitt F T, Moore R G, Johnston S, Cui Y-T, Li W, Yi M, Liu Z K, Hashimoto M, Zhang Y, Lu D H, Devereaux T P, Lee D-H and Shen Z-X 2014 *Nature* **515**, 245

59. Morris R C and Coleman R V 1973 *Phys. Rev. B* **7** 991-1001

60. Rusanov A Yu, Hesselberth M B S, Aarts J 2004 *Phys. Rev. B* **70** 024510

61. Maxfield B W and McLean W L 1965 *Phys. Rev.* **139**, A1515-A1522

62. Reyren N, Thiel S, Caviglia A D, Kourkoutis L F, Hammerl G, Richter C, Schneider C W, Kopp T, Rüetschi A-S, Jaccard D, Gabay M, Muller D A, Triscone J-M, Mannhart J 2007 *Science* **326**, 1196–1199

63. Logvenov G, Gozar A, Bozovic I 2009 *Science* **317**, 699–702

64. Di Castro D, et al 2012 *Phys. Rev. B* **86** 134524

65. Salvato M, et al 2013 *J. Phys.: Condens. Matter* **25** 335702

66. Ye J T, Zhang Y J, Akashi R, Bahramy M S, Arita R, Iwasa Y 2012 *Science* **388**, 1193-1196

67. Shi W, Ye J, Zhang Y, Suzuki R, Yoshida M, Miyazaki J, Inoue N, Saito Y and Iwasa Y 2015 *Scientific Reports* 5, 12534

68. Island J O *et al* 2016 *2D Materials* **3**, 031002

69. Costanzo D, Jo S, Berger H and Morpurgo A F 2016 *Nature Nanotechnology* **11**, 339-344

70. Xu C, Wang L, Liu Z, Chen L, Guo J, Kang N, Ma X-L, Cheng H-M and Ren W 2015 *Nature Materials* **14**, 1135- 1141

71. Cao Y *et al* 2015 *Nano Lett.* **15**, 4914-4921

72. Xi X et al 2016 *Nature Physics* **12**, 139-143

73. Yoshida M, Ye J, Nishizaki T, Kobayashi N Yoshihiro Iwasa Y 2016 *Appl. Phys. Lett.* **108**, 202602

74. Hess H F, Murray C A, and Waszczak J V 1992 *Phys. Rev. Lett.* **69**, 2138-2141

75. Saito Y, Kasahara Y, Ye J, Iwasa Y and Nojima T 2015 *Science* **350**, 409-413

76. Saito Y, Nojima T and Iwasa Y 2016 *Supercond. Sci. & Technol.* **29**, 093001

# On the origin of critical temperature enhancement in atomically-thin superconductors

*E.F. Talantsev[1,*], W.P. Crump[1], J.O. Island[2,**], Ying Xing[3,4,5], Yi Sun[3,4], Jian Wang[3,4], and J.L. Tallon[1,6]*

[1]Robinson Research Institute, Victoria University of Wellington,
69 Gracefield Road, Lower Hutt, 5040, New Zealand

[2]Kavli Institute of Nanoscience, Delft University of Technology,
Lorentzweg 1, Delft 2628 CJ, The Netherlands

[3]International Center for Quantum Materials, School of Physics, Peking University,
Beijing 100871, People's Republic of China

[4]Collaborative Innovation Center of Quantum Matter, Beijing, People's Republic of China

[5]Beijing Key Laboratory of Optical Detection Technology for Oil and Gas,
China University of Petroleum, Beijing 102249, People's Republic of China

[6]MacDiarmid Institute for Advanced Materials and Nanotechnology, P.O. Box 33436,
Lower Hutt 5046, New Zealand

*Corresponding author: evgeny.talantsev@vuw.ac.nz

**Present address: Department of Physics, University of California, Santa Barbara, CA 93106, USA

1. **Calculating error and mutual dependency of fitting parameters**

Following the method used by [1] we can estimate the variance of our calculated parameters in our model the following way:

The least squares method finds the model parameters by finding the minimum of the sum of squared residuals

$$SSE(p) = \sum(y_i - f(p, x_i))^2,$$

where the experimentally-observed dependent variable is $y_i$, the experimentally-observed independent variable is $x_i$, and $p$ is the set of parameters of the model $f(p, x_i)$.

The parameters $p$ will be optimised when

$$0 = \frac{\partial SSE(p)}{\partial p} = -2\sum(y_i - f(p, x_i))\frac{\partial f(p, x_i)}{\partial p}.$$

The estimated asymptotic covariance matrix of the estimated parameters can be formed:

$$\boldsymbol{V} = \sigma^2(\boldsymbol{F}^T\boldsymbol{F})^{-1},$$

where $F_{ij} = \frac{\partial f(\hat{p}, x_i)}{\partial p_j}$ and $\sigma^2 = \frac{SSE}{N-n_p}$ where $N$ is the total number of observations and $n_p$ is the number of fitting parameters. From the diagonal components matrix we find the estimated variance of each parameter ($\delta^2_{p_i} = V_{ii}$) and from the off diagonal components we find the estimated covariance between parameters ($\delta^2_{p_ip_j} = V_{ij}$).

We can also obtain an estimate of how much each parameter depends on the values of the others using the equation $dep_{p_i} = 1 - 1/V_{ii}(V^{-1})_{ii}$.

## 2. Calculating Adjusted $R^2$

We calculate the adjusted $R$-squared coefficient by using the following method:

$$R^2 = 1 - \frac{SSE}{SST}, SSE = \sum_i(y_i - f_i)^2, SST = \sum_i \sigma_i^2$$

$$\hat{R}^2 = 1 - (1 - R^2)\frac{N-1}{N-n_p-1}$$

Where $y_i$ is the experimentally-observed dependent variable, $f_i$ is the calculated dependent variable, and $\sigma_i$ is the estimated error for the variable $y_i$.

**References**

[1] J. Fox and H. S. Weisberg, "Nonlinear Regression and Nonlinear Least Squares in R," in *An R Companion to Applied Regression*, London, SAGE Publications Inc, 2011, p. Appendix.